# Adaptive Monitoring: A Systematic Mapping

Edith Zavala, Xavier Franch, Jordi Marco

zavala@essi.upc.edu, franch@essi.upc.edu, jmarco@cs.upc.edu

*Software and Service Engineering research group (GESSI),*
*Universitat Politècnica de Catalunya (UPC), Barcelona, Catalunya, Spain*

**Corresponding author:**

Edith Zavala

Software and Service Engineering research group (GESSI)

Universitat Politècnica de Catalunya (UPC)

Jordi Girona, 1-3, 08034

Barcelona, Catalunya, Spain

Tel: +34 644 20 22 23

Email: zavala@essi.upc.edu

Abstract

**Context**: Adaptive monitoring is a method used in a variety of domains for responding to changing conditions. It has been applied in different ways, from monitoring systems' customization to re-composition, in different application domains. However, to the best of our knowledge, there are no studies analyzing how adaptive monitoring differs or resembles among the existing approaches.

**Objective**: To characterize the current state of the art on adaptive monitoring, specifically to: a) identify the main concepts in the adaptive monitoring topic; b) determine the demographic characteristics of the studies published in this topic; c) identify how adaptive monitoring is conducted and evaluated by the different approaches; d) identify patterns in the approaches supporting adaptive monitoring.

**Method**: We have conducted a systematic mapping study of adaptive monitoring approaches following recommended practices. We have applied automatic search and snowballing sampling on different sources and used rigorous selection criteria to retrieve the final set of papers. Moreover, we have used an existing qualitative analysis method for extracting relevant data from studies. Finally, we have applied data mining techniques for identifying patterns in the solutions.

**Results**: We have evaluated 110 studies organized in 81 approaches that support adaptive monitoring. By analyzing them, we have: (1) surveyed related terms and definitions of adaptive monitoring and proposed a generic one; (2) visualized studies' demographic data and arranged the studies into approaches; (3) characterized the main approaches' contributions; (4) determined how approaches conduct the adaptation process and evaluate their solutions.

**Conclusions**: This cross-domain overview of the current state of the art on adaptive monitoring may be a solid and comprehensive baseline for researchers and practitioners in the field. Especially, it may help in identifying opportunities of research; for instance, the need of proposing generic and flexible software engineering solutions for supporting adaptive monitoring in a variety of systems.

1. **Introduction**

Over the years, methods and techniques for monitoring a variety of systems have been proposed. There are approaches proposed for monitoring communication networks (e.g., Liu et al. [1]), buildings' or persons' health (e.g., Kijewski-Correa et al. [2] and Mshali et al. [3], respectively), software systems (e.g., Toueir et al. [4]), environmental conditions (e.g., Alippi et al. [5]), etc. Monitoring allows systems' stakeholders checking how their systems progress or behave under different conditions, and reporting on relevant changes. However, it is often expensive and intrusive. Thus, the design of a monitoring system (i.e., the software system that implements monitoring capabilities) usually involves tradeoffs between the impact caused by the action of monitoring and its expected quality of results, such as data accuracy, freshness and coverage, among others [6,7]. In addition, a monitoring system is exposed to a diversity of runtime events, e.g., structural or operational changes on the System under Monitoring (SuM), faults on the monitoring system's elements or the emergence of new monitoring requirements.

In order to deal with all these challenging factors, software engineers have proposed different approaches for making current monitoring systems *adaptive*. Proposals have emerged from a variety of research fields (e.g., sensor networks, instrumentation, requirements monitoring). However, although these diverse proposals share most high-level challenges, solutions have been developed, evolved and kept isolated in those different fields. This hinders the discovery of synergies among the different proposals to adaptive monitoring as well as the standardization of the main field concepts and the normalization of the challenges faced. To the best of our knowledge, there is not any work reviewing adaptive monitoring approaches across different fields. Thus, this work aims at uncovering and characterizing existing approaches supporting the adaptation of monitoring systems, in general.

In order to achieve this goal, we have conducted a systematic mapping study for identifying the primary studies on adaptive monitoring published in academic venues. We have retrieved and selected the studies conducting a rigorous protocol, defined in this work, which follows the guidelines presented by Petersen et al. [8] and Kitchenham & Charters [9]. For analyzing the identified studies, we have designed 5 high-level research questions (RQs) which we have divided into a total of 18 research sub-questions. To extract data from these studies, we have used a qualitative analysis approach based on the method describe by Miles et al. [10]. After the qualitative analysis, we have applied data mining over the extracted data for identifying patterns in the approaches. Concretely, we have used the rule-based algorithm JRip implemented by the data mining tool Weka [11]. The results of this work not only provide an overview of the current state of the art on adaptive monitoring, but also improve the understanding about how the adaptation process is (usually) conducted by approaches proposed in different research fields. This is beneficial for facilitating, to the research community, the identification of reusable solutions, synergies, opportunities of improvement and unexplored methods and techniques.

The remainder of the paper is structured as follows. Section 2 presents background on adaptive monitoring and systematic mapping studies. Section 3 describes the process followed for conducting our systematic mapping. Section 4 presents the results of the systematic mapping and addresses the RQs. In Section 5, data mining techniques are applied to the data extracted in Section 4 and results are discussed. Finally, conclusions are presented in Section 6.

2. **Background**
2.1. **Adaptive monitoring**

Nowadays, the monitoring activity is integrated into control processes for gathering relevant data that is later analyzed by other software systems or the SuM administrators. The results of the analysis are mainly used for determining the state of the system and deciding whether any action (e.g., administering a medication when monitoring a person's health, or adapting a software service behavior in a nuclear plant) should be taken for keeping the SuM under control. Although some works consider the data gathering and analysis activities as part of a whole monitoring system (e.g., works by Bukenya et al. [12] and Ramirez et al. [7]), in this work we differentiate between them and focus on approaches that specifically support the adaption of the data gathering activity.

The adaptation of monitoring systems requires to manage and control their monitoring activity itself [13]. That is, monitoring systems' components and their operation should be supervised somehow as well, in order to determine monitoring systems' state and the adequacy of their data gathering strategies. According to Moui & Desprats [13], a monitoring strategy can be constructed by answering the questions: why do we monitor?, how do we monitor?, what do we monitor? and when do we monitor?. In our systematic mapping, we study how the state-of-the-art approaches analyze runtime data, and based on the analysis results, plan and execute the adaptation of the monitoring strategies as well as the monitoring systems' composition.

As it has been mentioned in Section 1, approaches for supporting adaptive monitoring have emerged from a variety of research fields. Thus, in this work the term "adaptive monitoring" becomes an abstract concept that is implemented in different ways by the different solutions. For instance, it can be implemented through the customization of a monitoring plan [14] or through the reconfiguration of a monitoring system's components [15]. One of the objectives of this work is to find or construct a definition for the term "adaptive monitoring" that could be applied in a generic way to all the approaches. We are also interested in uncovering other terms related to adaptive monitoring.

**2.2. Systematic mapping studies**

Systematic mapping studies (SMSs) or scoping studies are designed to give an overview of a research area through classification and counting contributions in relation to the categories of that classification [9,16]. It involves searching the literature in order to know what topics have been covered, and where the literature has been published [16]. SMSs share some commonalities with another type of empirical instrument, namely systematic literature reviews (e.g., with respect to searching and study selection). However, according to Petersen et al. [8], they are different in terms of goals and approaches to data analysis. While systematic literature reviews aim at synthesizing evidence, considering its strength, SMSs are primarily concerned with structuring a research area [8].

In order to ensure the quality of systematic reviews, a precise and rigorous methodology for conducting the review process has to be used. For this purpose, in this work, we follow the widely used guidelines proposed by Kitchenham & Charters [9] in conjunction with the updated ones for SMSs proposed by Petersen et al. [8]. The review process consists of three main phases:

- **Planning the review**. During this phase, all the decisions relevant to conducting the study are made. This includes the identification of the need for a review, the definition of the protocol for identifying primary studies and extracting the relevant data, and the definition of the visualization instruments and the validity threats of the study.
- **Conducting the review**: In this phase, the review process as defined during the planning phase has to be implemented. This process is iterative and may require revisions. It is recommended to record the information at all stages of the process.
- **Reporting the mapping**. Finally, this phase consists in reporting the results of the review. It includes specifying the dissemination mechanisms, the format of the report and the evaluation of the process.

In the rest of this work, we develop these phases for conducting our SMS on adaptive monitoring and report our results.

## 3. Planning the review

According to Petersen et al. [8] and Kitchenham & Charters [9], the planning phase of the review process consists of five main activities: need for a review identification and scoping, study identification, data extraction and classification, visualization and analysis of validity threats. In this section, we describe how we have performed each of these activities in our SMS. As recommended by Petersen et al. [8], some activities have been further split into sub-activities which are presented in separate subsections.

### 3.1. Need identification and scoping

The need identification and scoping activity has been divided in two sub-activities: need for a review identification and research questions definition. In the following subsections, we describe each of these sub-activities and detail how we have conducted them.

*3.1.1. Need for a review identification*

As stated by Petersen et al. [8], before carrying out any systematic literature study, researchers should identify and evaluate any existing systematic review on the topic of interest. Hence, in order to identify secondary studies on adaptive monitoring, we have followed a search protocol analogous to the main one presented in the study identification phase (see Section 3.2) of our SMS. In consequence, we have searched for existing reviews once the protocol was defined and before the SMS was conducted. In short, we have built a search string as a conjunction of population and intervention, as recommended by Kitchenham & Charters [9], and performed an automatic search on the databases of IEEE Xplore, ACM, Scopus and Inspect/Compendex

(Engineering Village). We have selected these databases based on the experience reported by Dybå et al. [17] and the results obtained by Petersen et al. [8] using them.

According to Kitchenham & Charters [9], in software engineering, the population may refer to a specific software engineering role, a category of software engineer, an application area or an industry group. In our context, the population corresponds to studies in the application area of adaptive monitoring (see Table 1). On the other hand, they refer to the intervention as a software methodology/tool/technology/procedure that addresses a specific issue. In our case, the intervention is systematic mappings (see Table 1). In order to increase the number of results, from each main term, we have defined a set of synonyms, variants and acronyms (see Table 1). Wildcards have not been used because: 1) some databases do not support the number of wildcards per search we would require; 2) in this way, we dramatically reduce the number of noisy studies. We have constructed the search string by applying the Boolean OR operator to link the Population terms and Intervention terms presented in Table 1 separately, and a Boolean AND operator to link these two resulting substrings.

**Table 1**
Search string terms.

| Type | Main term | Alternative terms (Synonyms/Variants/Acronyms) | |
| --- | --- | --- | --- |
| Population | Adaptive monitoring | adaptive monitor<br>adaptive monitors<br>adaptable monitoring<br>adaptable monitor<br>monitor adaptation<br>monitoring adaptation<br>reconfigurable monitor<br>reconfigurable monitoring<br>monitoring reconfiguration<br>dynamic monitor<br>dynamic monitors<br>dynamic monitoring<br>monitoring evolution<br>monitor evolution<br>monitors evolution | evolving monitoring<br>evolutionary monitoring<br>monitoring customization<br>customized monitor<br>customized monitors<br>customized monitoring<br>customised monitoring<br>monitoring personalization<br>personalized monitors<br>personalized monitoring<br>personalised monitoring<br>reactive monitoring<br>reactive monitors<br>proactive monitoring |
| Intervention | Systematic mappings | systematic mapping<br>state of the art<br>SLR<br>review | |

The search resulted in 271 papers. Then, we have applied a study selection protocol similar to the one applied in our SMS. The only difference is the inclusion/exclusion criteria we have used for selecting the studies of interest. In this case, the inclusion criteria that have been applied were:

- Studies present summaries of adaptive monitoring approaches.
- Studies are in the fields of computer science or engineering.
- Studies were published until 2016.

For excluding studies, we have applied the following criteria:

- Studies present non-peer reviewed material.
- Studies not written in English.
- Studies not accessible in full-text.

After applying the selection protocol, we have not found any secondary study on the adaptive monitoring topic, neither in general nor in a particular research field. However, when performing the snowballing process in our SMS, we have been able to identify one related work [18]. Although this work is not focused on adaptation and surveys only approaches supporting energy-efficient wireless sensor networks, we have considered it worth to mention since it has been the only review we have found related to our work. As we will explain later in Section 3.2, the approaches cited in this survey that provide energy-conservation through the adaptation of the data gathering activity have been considered in our SMS.

*3.1.2. Research questions definition*

Given this lack of secondary studies, we consider that conducting a SMS in the adaptive monitoring topic is important and justified. The SMS we conduct in this work aims at giving a comprehensive overview of the current state of the art of the adaptive monitoring topic and improving the understanding about how approaches (tend to) conduct the adaptation process. In order to reach this goal, we have designed 5 high-level RQs (see Table 2). The RQs are exploratory, as we attempt to understand the current state of the adaptive monitoring topic and identify key aspects about how it is supported by the different existing approaches. The RQs have been divided into a total of 18 research sub-questions, as it is shown in Table 2.

**Table 2**

Research questions of the review.

| Research Question | Sub-question | |
|---|---|---|
| RQ1. What is adaptive monitoring? | RQ1.1 | What are the terms related to the term "adaptive monitoring"? |
| | RQ1.2 | Are there specific definitions of adaptive monitoring? |
| | RQ2.1 | When are the studies published? |
| RQ2. What are the demographic characteristics of the studies about adaptive | RQ2.2 | Where are the studies published? |
| | RQ2.3 | How are publications distributed between academy and industry? |
| | RQ2.4 | How publications are geographically distributed? |
| monitoring? | RQ2.5 | How are the studies organized into approaches for adaptive monitoring? |
| RQ3. What is proposed by adaptive monitoring approaches? | RQ3.1 | What type of contributions is presented? |
| | RQ3.2 | How generic are the solutions presented? |
| RQ4. How adaptive monitoring is conducted by the approaches? | RQ4.1 | What is the purpose of adaptation? |
| | RQ4.2 | What is adapted? |
| | RQ4.3 | What triggers adaptation? |
| | RQ4.4 | How analysis is performed? |
| | RQ4.5 | How adaptation decisions are made? |
| | RQ4.6 | How adaptation decisions are enacted in the monitoring system? |
| | RQ4.7 | What type of adaptation is executed? |
| RQ5. How adaptive monitoring approaches are evaluated? | RQ5.1 | What type of evaluation is performed? |
| | RQ5.2 | In which type of systems is the evaluation performed? |

### 3.2. Study identification

The study identification activity has been divided into three sub-activities: search string construction, literature sources identification and study selection.

*3.2.1. Search string construction*

The aim of the search process in our SMS is to find as many primary studies related to the RQs as possible using an unbiased search strategy. In order to build the search string, we have followed again the recommendation of Kitchenham & Charters [9] and created the string as a conjunction of population and intervention. As in the previous search, our population is composed by studies in the application area of adaptive monitoring. What has changed in this search is the intervention: we are now interested in approaches supporting adaptive monitoring and not in SMSs. In order to increase the number of results, as we have previously done in our first search in Section 3.1.1, we have defined a set of synonyms and variants for the main search terms (i.e., adaptive monitoring and approaches). In the case of the population, we have reused the alternative terms identified in Section 3.1.1 (see Table 1).

While evaluating the articles resulting from the search of Section 3.1.1, we have noticed that the *dynamic monitor, dynamic monitors* and *dynamic monitoring* terms, included in the search string, have been utilized by some of the studies for referring to the continuous runtime monitoring of dynamic factors (e.g., as it has been used by Bukenya et al. [12] or Magalhães et al. [19]) or as an adjective to describe how the adaptation process is actually conducted (e.g., as it has been used by Clark et al. [20] or Jeswani et al. [21]). Thus, in order to avoid noisy papers, we have decided not to consider these terms for the search string of our SMS. Regarding the intervention, we have identified the alternative terms: approach, method, framework and technique. The search string has been constructed using the terms and the Boolean OR and AND operators as we have done in Section 3.1.1.

*3.2.2. Literature resources identification*

In order to identify primary studies, researchers can perform either automatic search through the usage of scientific databases or manual search through gathering the studies from specific known journals and conferences of the target field. Both approaches present advantages and drawbacks. The most common way of searching is the automatic search, followed by the manual search [8]. However, in this work, this was not possible since we were not able to identify any relevant dedicated conference or journal in the specific field of adaptive monitoring (neither before nor after conducting the data extraction process). For this reason, similarly to Petersen et al. [8], we have decided to conduct an automatic search and complement it with a backward snowball sampling of all studies selected after full-text reading.

In order to select the databases for conducting the automatic search, we followed the same criteria as in Section 3.1.1, since we have not found any other secondary study in the adaptive monitoring topic for guiding the search. Thus, the databases used in our SMS are IEEE Xplore, ACM, Scopus and Inspect/Compendex (Engineering Village). As recommended by Petersen et al. [8], we have used a tool for managing the references extracted from the databases and a tool for recording extracted data. Concretely, we have used the reference management tool Mendeley and the qualitative data analysis tool Atlas.ti® ([www.atlasti.com](www.atlasti.com)).

*3.2.3. Study selection*

In order to select the final set of studies, we have designed a study selection strategy similar to the one presented by Franco-Bedoya et al. [22]. Concretely, the strategy consists of a set of stages, shown in Fig. 1, and is an adaptation of the steps proposed by Petersen et al. [8] and Kitchenham & Charters [9]. In Fig. 1, we provide an overview of the study selection process and the number of papers resulting in each stage. Fig. 1 also details the backward snowballing process we have conducted in the last stage of our study selection strategy.

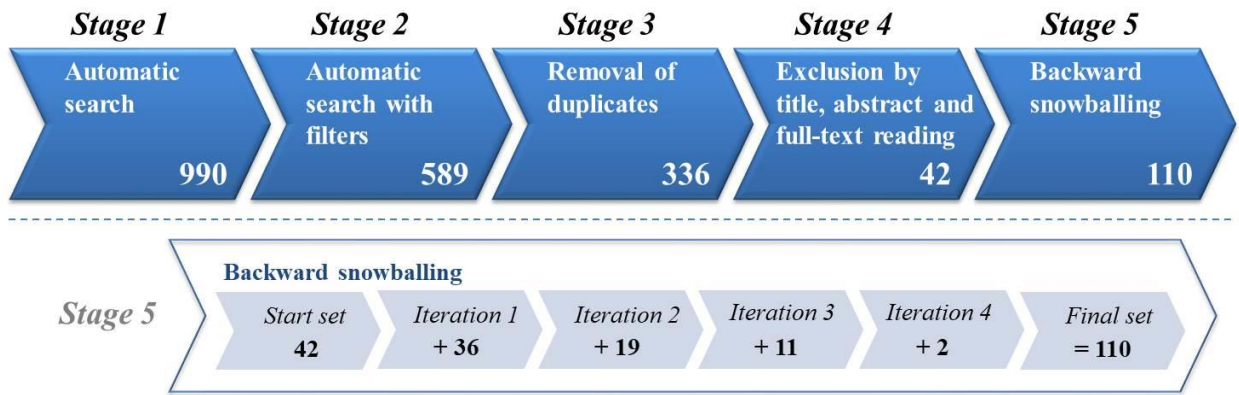

Figure 1: Study selection strategy

The exclusion of studies has been done based on titles and abstracts, as well as full-text reading. In order to identify as many primary studies as possible, we have also added studies through backward snowballing. The application of the inclusion and exclusion criteria has been conducted by the first author. Along the process, periodical meetings have been held with the rest of the authors for discussing and refining the final set of included and excluded papers. The following inclusion criteria have been applied to the studies:

- Studies present a solution (i.e., approach, method, framework, technique or others) for supporting adaptive monitoring.
- Studies are in the fields of computer science or engineering.
- Studies were published until 2016.

Studies fulfilling the following criteria have been excluded:

- Studies are secondary studies.
- Studies present work in progress.
- Studies present non-peer reviewed material.
- Studies are not written in English.
- Studies are not accessible in full-text.
- Studies are books, books reviews or grey literature.

In the rest of this section, we provide the details of each of the stages of the study selection strategy shown in Fig. 1.

- **Stage 1. Automatic search.** This stage corresponds to the automatic search on the digital databases we have detailed in Section 3.2.2. As a result of this stage, 990 primary studies have been identified. Table 3 shows how many studies have been extracted per database (see column *Search results*).
- **Stage 2. Automatic search with filters.** After performing the automatic search, we have applied a set of filters that some of the digital libraries offer for excluding studies that are not of our interest. The filters correspond to some of the inclusion/exclusion criteria we have listed before in this section. Table 3 shows the filters we have used in each database and the resulting number of articles after applying those filters (see column *Filtered search results*). As a result of this stage, 401 papers have been automatically discarded, resulting in 589 primary studies.
- **Stage 3. Removal of duplicates.** From the 589 papers identified in the previous stage, we have automatically removed duplicated studies by using the reference manager Mendeley. In addition, the first author has manually reviewed the list of articles in order to identify duplicated records (no detected by Mendeley). As a result, 253 articles have been excluded. That is, after this stage we have ended up with 336 remaining primary studies.

- **Stage 4. Exclusion by title, abstract and full-text reading.** In this stage, the first author has reviewed all the titles and abstracts and applied the inclusion and exclusion criteria for each study. A paper has been taken to full-text reading when in doubt and discussed with the rest of the authors. The final set of included and excluded papers has been revised through a series of periodic meetings involving all the authors. After this stage, 294 out of the 336 studies resulting from the previous stage have been excluded, resulting in 42 remaining articles.

**Table 3**
Number of studies per database with filters applied.

| Database | Filters | Search results | Filtered search results |
| --- | --- | --- | --- |
| **IEEE** | No filters applied | 95 | 95 |
| **ACM** | *Exclude*: 2017 | 85 | 84 |
| **Scopus** | *Exclude*: 2017<br>*Limit to:*<br>– *Subject Area*: Computer Science, Engineering<br>– *Document Type*: Conference paper, Article<br>– *Language*: English | 440 | 238 |
| **Inspect/ Compendex** | *Exclude:* 2017<br>*Limit to:*<br>– *Classification code*: Computer Software, Data Handling and Applications, Computer Applications, Control Systems, Digital Computers and Systems, Computer Systems and Equipment, Automatic Control Principles and Applications, Distributed Systems Software, Software Engineering techniques<br>– *Document type*: Conference article, Journal article, Conference proceeding<br>– *Language*: English | 370 | 172 |

- **Stage 5. Backward snowballing.** In order to identify as many primary studies as possible, we have conducted a backward snowballing process organized into 4 iterations (see Fig. 1). The process' start set has been composed of the articles that have resulted from Stage 4 (42). While iterating, relevant works have been identified from the reference list of the articles. During the first iteration of the snowballing process, we have identified a secondary study relevant to our SMS [18]. As we have explained in Section 3.1.1, this secondary study surveys approaches supporting energy-efficient wireless sensor networks. Due to the inclusion and exclusion criteria, this study has not been included in our final set of articles. However, since we have identified that some of the surveyed approaches' solutions involve the adaptation of the data gathering activity, we have taken this secondary study into account when performing the backward snowballing process. That is, for the second iteration we have included the relevant works identified in the reference list of this secondary study.

The referenced works have been included based on the inclusion and exclusion criteria we have previously defined in this section. Moreover, we have decided to exclude papers published before 2000 (publication year of the oldest start set paper is 2001). Fig. 1 shows the number of papers we have extracted during the process and that fulfill the inclusion criteria (i.e., the secondary study mentioned before, identified in iteration 1, has been omitted in the image for the sake of simplicity). As recommended by Wohlin (2014), we have finished the snowballing when no new papers fulfilling our criteria have been found. As a result of this stage, 68 papers have been added to the start set, resulting in a final set of 110 relevant primary studies for our SMS.

The final set of primary studies is further analyzed in the following sections.

### 3.3. Data extraction and classification

In order to extract the data from the primary studies, we have used a qualitative data analysis approach based on the method described by Miles et al. [10]. The qualitative data analysis tool Atlas.ti® has been used for supporting this process and ensuring consistent and accurate extraction of the key information related to the RQs. The extraction process has been performed by the first author and reviewed and confirmed by the other two authors. Extracted data has been discussed by all the authors in a series of periodic meetings scheduled for this purpose. To extract data from the primary studies, we have developed the template shown in Table 4. The qualitative analysis has consisted of the following three main steps:

- **Data extraction preparation.** In this step, the 110 primary studies included in our SMS have been imported into a new Atlas.ti® project.
- **First cycle coding.** Codes are labels that assign symbolic meaning to the descriptive or inferential information compiled during a study. They are primarily, but not exclusively, used to retrieve and categorize similar data chunks so the researcher can quickly find, pull out, and cluster the segments relating to a particular RQ, hypothesis, construct, or theme [10]. In order to create the codes of our SMS, we have performed both deductive and inductive coding. First, we have defined a start list of codes from the RQs, i.e., deductive coding. Then, we have added codes that progressively emerged during the data extraction process, i.e., inductive coding. Table 4 shows the information extracted from the primary studies (i.e., data extraction forms) that we have used to define the codes.
- **Second cycle coding (pattern codes).** In this step, codes have been grouped into smaller number of categories, themes, or constructs (i.e., pattern codes). Pattern codes are explanatory or inferential codes that identify an emergent theme, configuration, or explanation [10]. In Section 4, the pattern codes of this SMS are described in the RQs where they have been identified.

The process has consisted of several iterations in which codes were added, modified and removed over time in order to ensure the validity and consistency of the results.

**Table 4**
Data extracted from primary studies

| Data item |
| --- |
| Full reference. |
| Year of publication. |
| Source (conference, journal, workshop). |
| Type of publication (academy, industry). |
| First author's affiliation (organization and country). |
| Relation(s) with other primary studies of this SMS (references, references and extends, extends). |
| Term(s) used for referring to the data gathering activity adaptation. |
| Definition(s), if any, of adaptive monitoring. |
| Type of main research contributions (algorithm(s), architecture) of the approach and its generalizability level (problem-specific, domain-specific, generic). |
| Approach purpose of adapting the monitoring system. |
| Monitoring system's element(s) adaptation supported by the approach. |
| Approach adaptation process trigger(s). |
| Method(s), if any, used by the approach for analyzing relevant runtime data. |
| Method(s) used by the approach for (planning and) making the adaptation decision(s). |
| Type of adaptation decision enactment process supported by the approach (manual, semi-automatic, automatic). |
| Type of adaptation executed by the approach for adapting the monitoring system (structural, parameter). |
| Type of approach evaluation (experiment, industry use case), if any, and type of system in which the evaluation is performed. |

### 3.4. Visualization

In order to present the findings of our study, we have used different kind of methods (e.g., tables and charts) (see Section 4). The goal is to condense the major data for further analysis and to represent and present the conclusions. Table 5 presents the variables that have been tabulated and are used to answer the RQs.

**Table 5**
Data tabulated per research question

| Data | RQ |
| --- | --- |
| Terms related to adaptive monitoring | RQ1.1 |
| Number of studies per term related to adaptive monitoring | RQ1.1 |
| Year at which each term related to adaptive monitoring has been first and last used | RQ1.1 |
| Sources of adaptive monitoring definitions | RQ1.2 |
| Adaptive monitoring definitions | RQ1.2 |
| Number of studies per year | RQ2.1 |
| Number and percentage of studies per type of source and year | RQ2.2 |
| Number and percentage of studies per type of publication and year | RQ2.3 |
| Number and percentage of studies per continent and year | RQ2.4 |
| Number of studies per country | RQ2.4 |
| Studies per approach and research field | RQ2.5 |
| Studies citation relation(s) with other studies of this SMS | RQ2.5 |
| Number and percentage of approaches per type of contribution and year | RQ3.1 |
| Number and percentage of approaches per generalizability level and year | RQ3.2 |
| Number and percentage of approaches per type of adaptation purpose and year | RQ4.1 |
| Number of approaches per combination of types of adaptation purposes (for most relevant combinations) | RQ4.1 |
| Number of approaches per adaptation purpose (for most relevant types) | RQ4.1 |
| Number and percentage of approaches per element adapted and year | RQ4.2 |
| Number of approaches per combination of elements adapted (for most relevant combinations) | RQ4.2 |
| Number and percentage of approaches per type of adaptation trigger and year | RQ4.3 |
| Number of approaches per combination of types of triggers (for most relevant combinations) | RQ4.3 |
| Number of approaches per adaptation trigger (for most important types) | RQ4.3 |
| Number and percentage of approaches per analysis method and year | RQ4.4 |
| Number of approaches per combination of analysis methods (for most relevant combinations) | RQ4.4 |
| Number and percentage of approaches per decision-making method and year | RQ4.5 |
| Number of approaches per combination of decision-making methods (for most relevant combinations) | RQ4.5 |
| Number and percentage of approaches per type of enactment and year | RQ4.6 |
| Number and percentage of approaches per type of adaptation executed and year | RQ4.7 |
| Number and percentage of approaches per type of evaluation and year | RQ5.1 |
| Number and percentage of approaches per type of system in which the evaluation is performed and year | RQ5.2 |

### 3.5. Validity threats

For any empirical study the discussion of validity threats is of importance and is a quality criterion for study selection [8]. This section presents the aspects of the research process that might represent threats to validity and the actions performed to mitigate them. According to the recommendations by Petersen et al. [8], the types of validity threats that should be taken into account are: descriptive validity, theoretical validity, generalizability validity, interpretive validity and repeatability.

*3.5.1. Descriptive validity*

Descriptive validity is the extent to which observations are described accurately and objectively [8]. In order to reduce this threat, we have designed a data extraction template for supporting the recording of data. The template tries to objectify the data extraction process. The different template items, in the form of codes, are linked to specific parts of the primary studies, so they can be revisited when required, as it has been the case during the analysis.

*3.5.2. Theoretical validity*

Theoretical validity is determined by our ability of being able to capture what we intend to capture. Confounding factors such as biases and selection of subjects play an important role [8]. In order to reduce this threat, first, in the study identification process we have complemented the automatic search with backward snowballing of all studies. Then, since the selection process has mainly been conducted by the first author (biases may appear), we have scheduled a set of periodic meetings with the rest of the authors for discussing and refining the final set of included and excluded papers.

This study has been conducted during 2017 and written during end of 2017 and beginning of 2018. Hence, only studies from 2016 and earlier have been included in the analysis. In spite of this limitation, we consider our sample of primary studies a good representation since a total of 110 studies, organized in different approaches proposed from different monitoring application domains, were identified (see Fig. 8). Furthermore, different types of publication venues are well represented (see Fig. 4). Finally, during the extraction process, codes have been created by the first author what could also affect the validity of this task. In order to reduce this threat, the second and third authors have assessed the extracted data. Though, given that this step involves human judgment, the threat cannot be eliminated [8].

*3.5.3. Generalizability validity*

There are two types of generalizability validity, internal and external [8]. Given that the identified primary studies come from different monitoring application domains and research fields, we consider internal generalizability not a major threat of this SMS. Regarding the external generalizability, since the results of our SMS are within the scope of adaptive monitoring and we do not attempt to generalize conclusions beyond this scope, validity threats in this regard do not apply.

*3.5.4. Interpretive validity*

Interpretive validity is achieved when the conclusions drawn are reasonable given the data, and hence maps to conclusion validity [8]. In order to reduce this threat, the experienced second and third authors have revised insights obtained by the first author and discussed with her possible misunderstandings.

*3.5.5. Repeatability*

The repeatability requires detailed reporting of the research process [8]. We have reported the process we have followed for conducting our SMS, and also described the actions taken to reduce threats to validity. We have also helped repeatability by using existing guidelines for conducting the review.

## 4. Results of the review

In this section, we address the RQs introduced in Table 2. With this goal, we summarize the results obtained from the data extraction process described in Section 3.3. The data extracted from the studies and used to address the RQs is available at [24].

### 4.1. RQ1. What is adaptive monitoring?

*RQ1.1 - What are the terms related to the term "adaptive monitoring"?*

In order to answer this question, we have performed a first cycle coding (see Section 3.3) using the In Vivo method defined by Miles et al. [10]. Next, we have categorized those codes into different groups. We were looking for other terms used by the researchers for referring to adaptive monitoring. We have found that 90 out of the 110 studies use other terms for referring to adaptive monitoring. We have grouped these different terms into 33 categories. In order to do that, we have identified terms that could be variants of a simpler term and put them into the same group (e.g., Monitoring Reconfiguration, Reconfigurable Monitors, Self-configuring Monitoring, Monitors Configuration, among other similar terms, can been grouped into a category named Monitor Configuration). Some of the terms could not be grouped with others; therefore, a category for each of them has been created.

In Fig. 2, we present the most relevant categories (i.e., categories of terms mentioned in more than one study) ordered by the total number of studies that mentioned them (number in parenthesis). Categories composed by more than one term are marked with an asterisk. We have included a category named *Adaptive Monitor* in which we grouped terms such *Adaptive Monitoring* or *Adaptive Monitors*. Terms in the *Adaptive Monitor* category have been found in 41 out of the 110 studies of this SMS. Regarding the rest of terms, as it can be noticed, terms grouped into the *Monitor Configuration* category are the most mentioned (17 papers) followed by *Adaptive Sampling* (16 papers).

In this RQ, we were also interested on studying the way in which these terms have been used over the years. Thus, for each category, we have determined the year at which its terms have been first and last used. Fig. 2 shows how some groups of terms are well-established in the community with a long life span (e.g., *Adaptive Monitor*, *Monitor Configuration* and *Adaptive Sampling*) while others show some obsolescence (e.g., *Active Probing*) or even a spurious momentum (e.g. *Conditional Data Acquisition*).

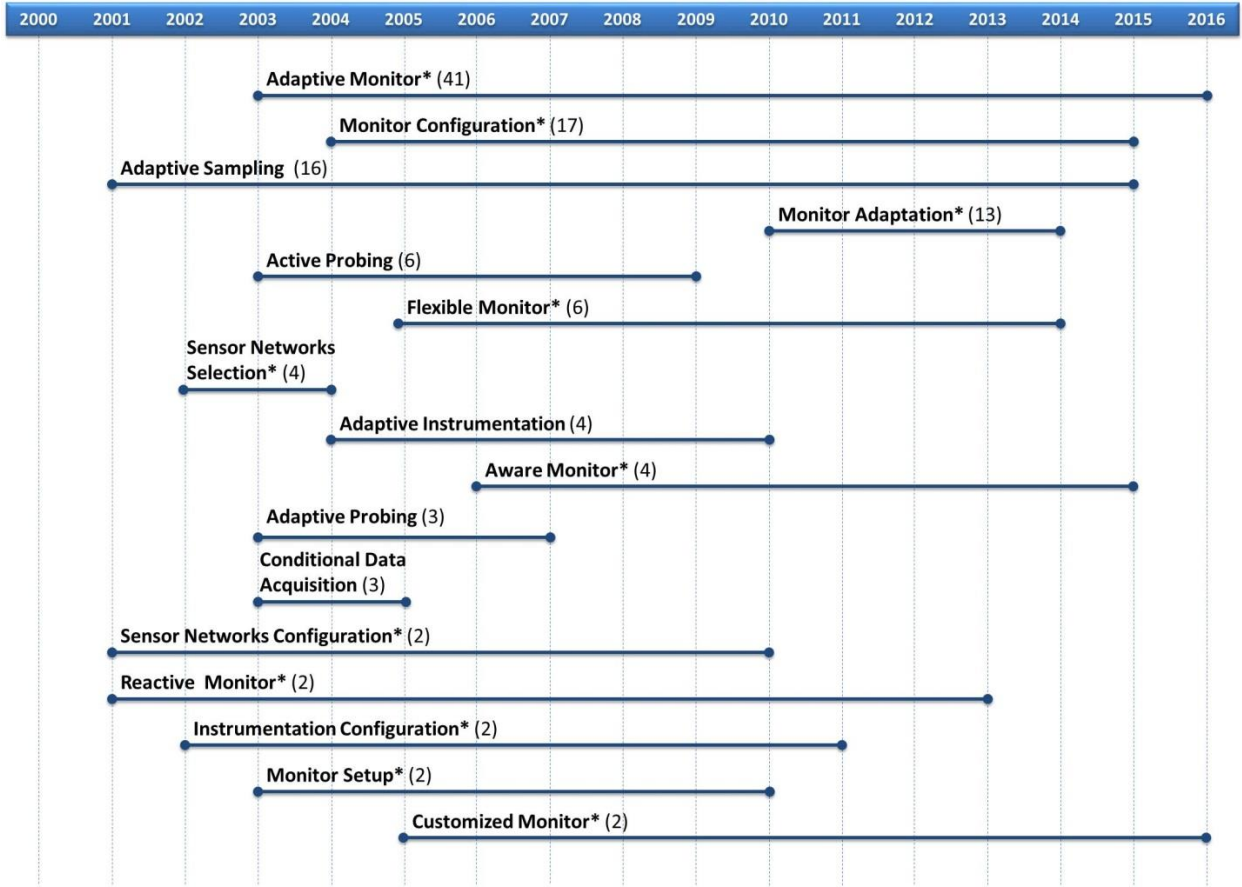

Figure 2: Categories of terms related to the term "adaptive monitoring" present in more than one study over the years

*RQ1.2 - Are there specific definitions of adaptive monitoring?*

For answering this RQ, we have applied first cycle coding (see Section 3.3), restricted to the *Adaptive Monitor* category introduced in RQ1.1. As a result, we have identified that in the majority of the studies, there was no interest on defining the terms in the Adaptive Monitor category but instead on describing how they are actually realized (e.g., adjusting a variable, reconfiguring components). Specifically, we have only found 2 out of the 41 studies that actually present a definition for the term Adaptive Monitoring. Both works are from the same authors and the definition presented was the same as well. Concretely, authors define Adaptive Monitoring as:

> *"The ability an online monitoring function has to decide and to enforce, without disruption, the adjustment of its behavior for maintaining its effectiveness, in respect of the variations of both functional requirements and operational constraints, and possibly for improving its efficiency according to self-optimization objectives."* [25,26]

## 4.2. RQ2. What are the demographic characteristics of the studies about adaptive monitoring?

*RQ2.1 - When are the studies published?*

To answer this RQ, we have also applied only first cycle coding (see Section 3.3). Concretely, we have created a pre-defined list of codes deduced from the publication years we are considering in our systematic mapping (2000 to 2016). Fig. 3 shows the number of studies published per year.

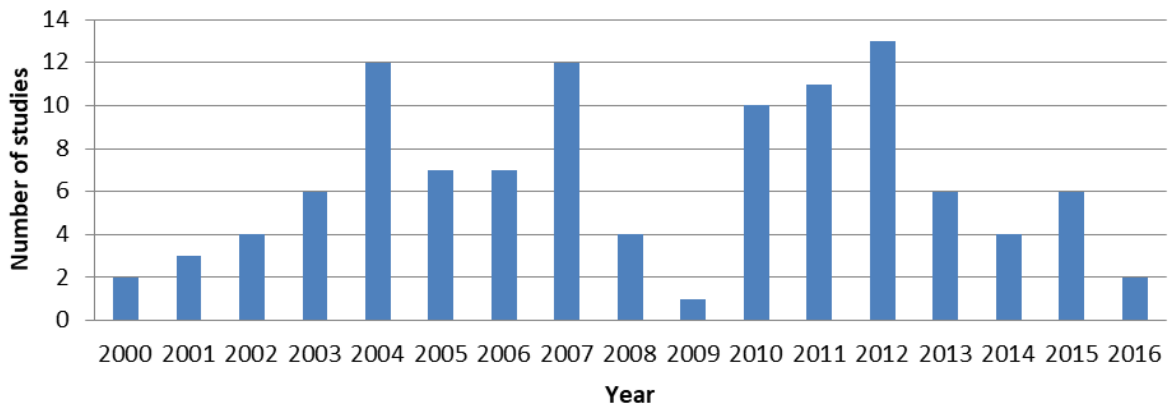

Figure 3: Number of studies published per year

*RQ2.2 - Where are the studies published?*

For addressing this RQ, we have conducted an inductive first cycle coding using the In Vivo [10] method on the name of the sources. Then, we have classified the sources by type: *Conference*, *Journal* and *Workshop*. The distribution of the 110 primary studies among these categories is shown in Fig. 4a. According to our data, conference proceedings (with 68 papers) are the most prevalent publication type. Fig. 4b shows the percentage of studies published in the different types of sources per year.

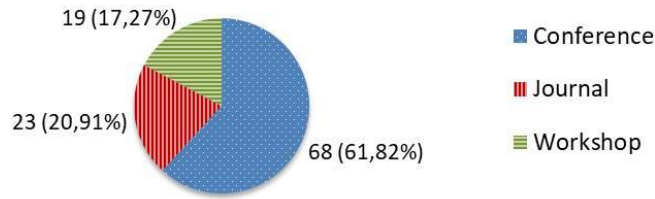

(a)

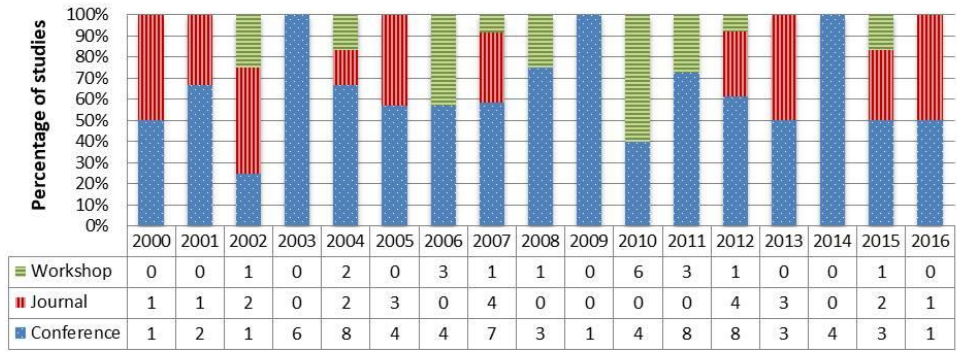

(b)

Figure 4: Number and percentage of studies per source type: (a) total, (b) over the years.

*RQ2.3 - How are publications distributed between academy and industry?*

In order to answer this question, we have analyzed whether at least one of the authors in each study came from a non-academic institution (similarly to the approach applied by Franco-Bedoya et al. (2017)) and applied first cycle coding. Fig. 5a shows that 26 out of the 110 primary studies are from *Industry*, while 84 studies are from *Academy*. In Fig. 5b, we provide an overview of the percentage of *Industry* and *Academy* studies published per year.

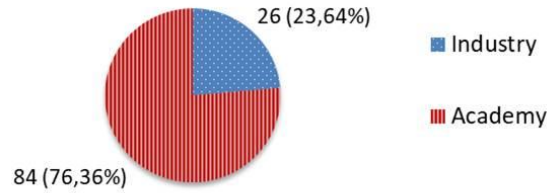

(a)

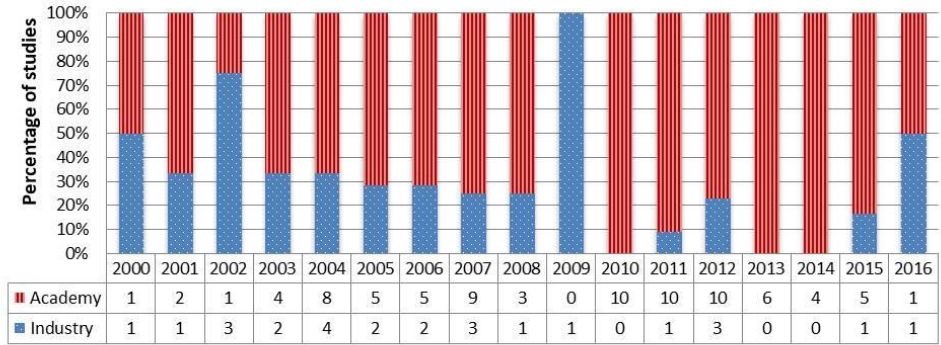

(b)

Figure 5: Number and percentage of industry and academy studies: (a) total, (b) over the years.

*RQ2.4 - How are publications geographically distributed?*

For addressing this RQ, we have conducted a first cycle coding using the In Vivo [10] method on the whole affiliation information of the first author of each study. Then for the second cycle of coding, we have done two iterations: first, we have categorized affiliations per country; second, we have grouped countries by continents. In Fig. 6a, we show the distribution of studies among the different continents. *North America* (57 papers) and *Europe* (41 papers) are the most dominant continents. Fig. 6b provides information about the percentage of studies published in each continent by year. It can be noticed that until 2010, studies were mainly published by institutions placed in *North America*. Afterwards, *Europe* takes the lead. Finally, in Fig. 7 we display how studies are geographically distributed among the different countries. *USA* is by far the country with more published studies (51 papers).

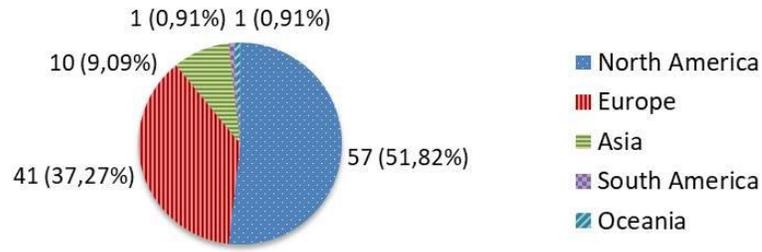

(a)

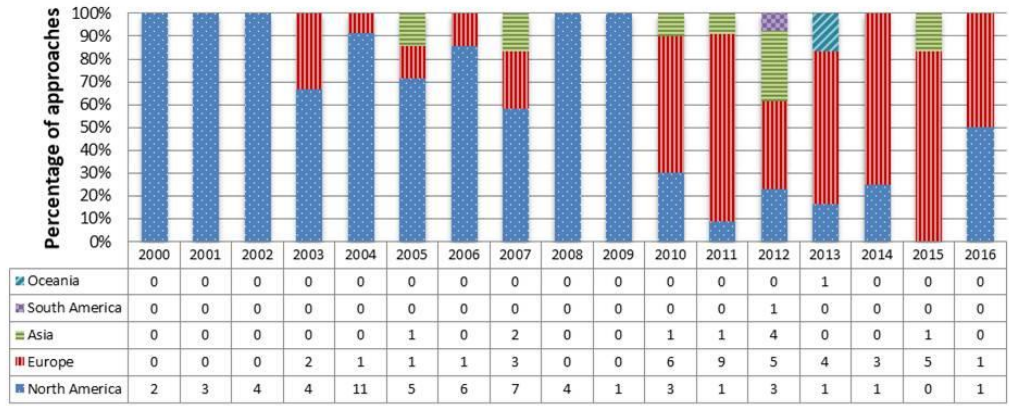

(b)

Figure 6: Number and percentage of studies published per continent: (a) total, (b) over the years.

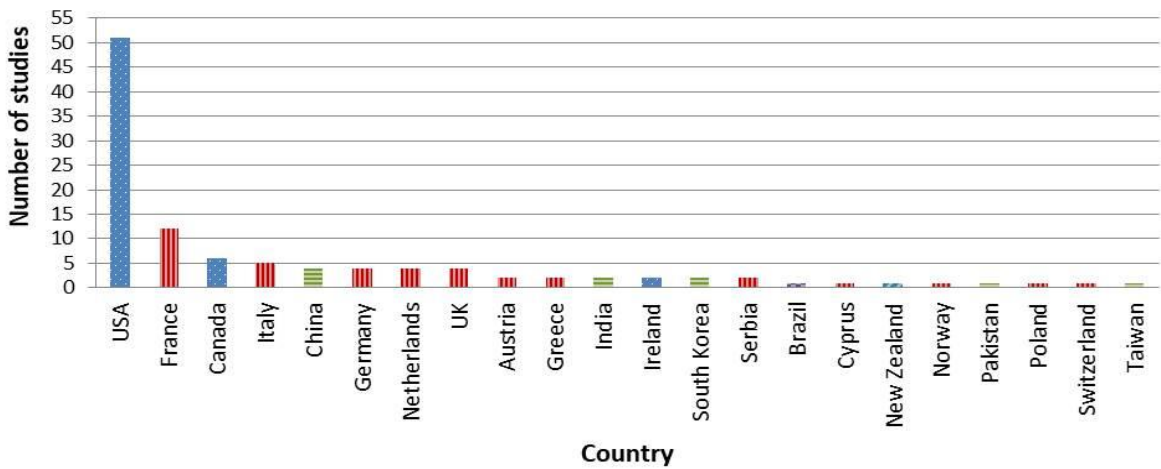

Figure 7: Number of studies published per country

*RQ2.5 – How are the studies organized into approaches for adaptive monitoring?*

In order to organize the studies into approaches, we have determined, based on the list of authors and full-text reading of the articles, which studies were extended by other studies (i.e., belong to the same approach according to our interpretation). We have conducting a first cycle coding, creating a network of the 110 primary studies, using Atlas.ti® in which we indicate which studies reference and extend, or are extended by (but not referenced by), other studies. As a result, 81 approaches have been identified, 64 composed of only one study and 17 consisting of more than one. In Fig. 8, we represent the 110 studies by small circles. We have assigned to each circle a resource identifier (extracted from the list of references provided in Table A1, Appendix A). The studies that are part of the same approach have been grouped into bigger circles (circles numbered from 1 to 17 in Fig. 8).

During the analysis, we have also extracted the citation information among the studies of our systematic mapping. In Fig. 8, this information is shown in the form of arrows. Some of the studies had not citation relation with other studies of the systematic mapping. In Fig. 8 these studies were grouped into the rectangle placed at the bottom of the figure. Once placed, we have further classified the studies in different abstract topics or research fields that predominated on each cluster. The categories are shown in Fig. 8 in the form of circles tagged with the topic or field name. The rectangle containing the studies without citation relation has been tagged as *Various* (since those studies are cross-domain).

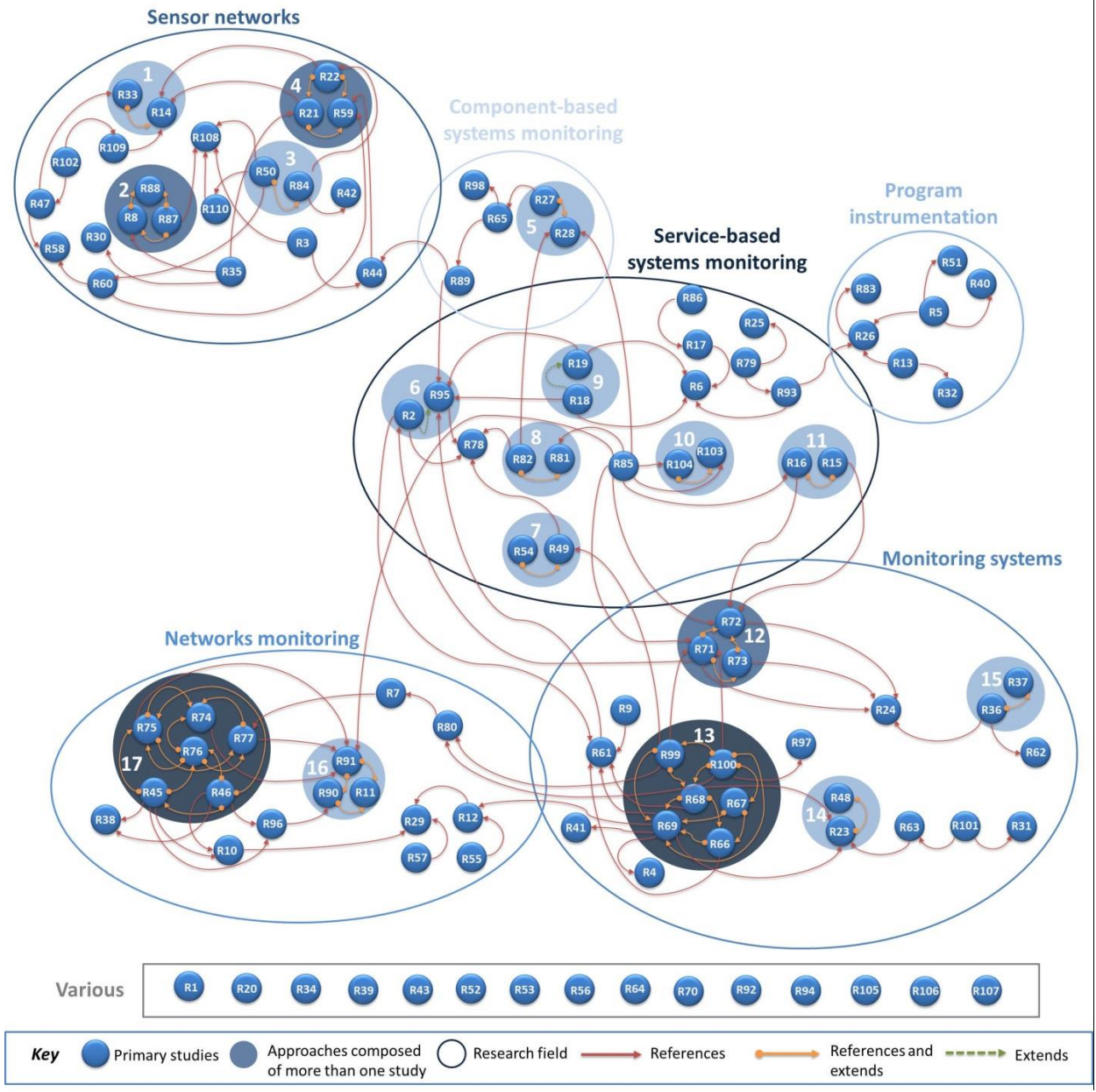

Figure 8: Studies organized by citing information in approaches and research fields

**4.3. RQ3. What is proposed by adaptive monitoring approaches?**

Given their more profound intent, the research questions RQ3, RQ4 and RQ5, have been analyzed considering the 81 approaches instead of the individual papers. For the 17 approaches composed by more than one study, we have mainly based the analysis on either the latest published study of the set or the most complete version (e.g., journal publications may provide more details than conference proceedings). Occasionally, we have revised other studies of the set to clarify unclear issues. It is worth remarking that, when visualizing the approaches by year, we have used the year of the last contribution, i.e., the study of the set with the latest publication date. Finally, in order to focus on trends when further exploring second cycle categories (when applicable), we calculate the average number of approaches per category in each research sub-question and focus on categories present in a total number of approaches above this average.

*RQ3.1 - What type of contributions are presented?*

Based on the type of proposals presented by the studied approaches, we have derived the codes: *Algorithm(s)-only* and *Algorithm(s) and architecture* with which we have conducted a first cycle coding (see Section 3.3). Fig. 9a shows that the contributions of 42 approaches are of the type *Algorithm(s) and architecture* while the contributions of 39 approaches are *Algorithm(s)-only*. In Fig. 9b, we condense the information about the percentage of published approaches per type of contribution over the years.

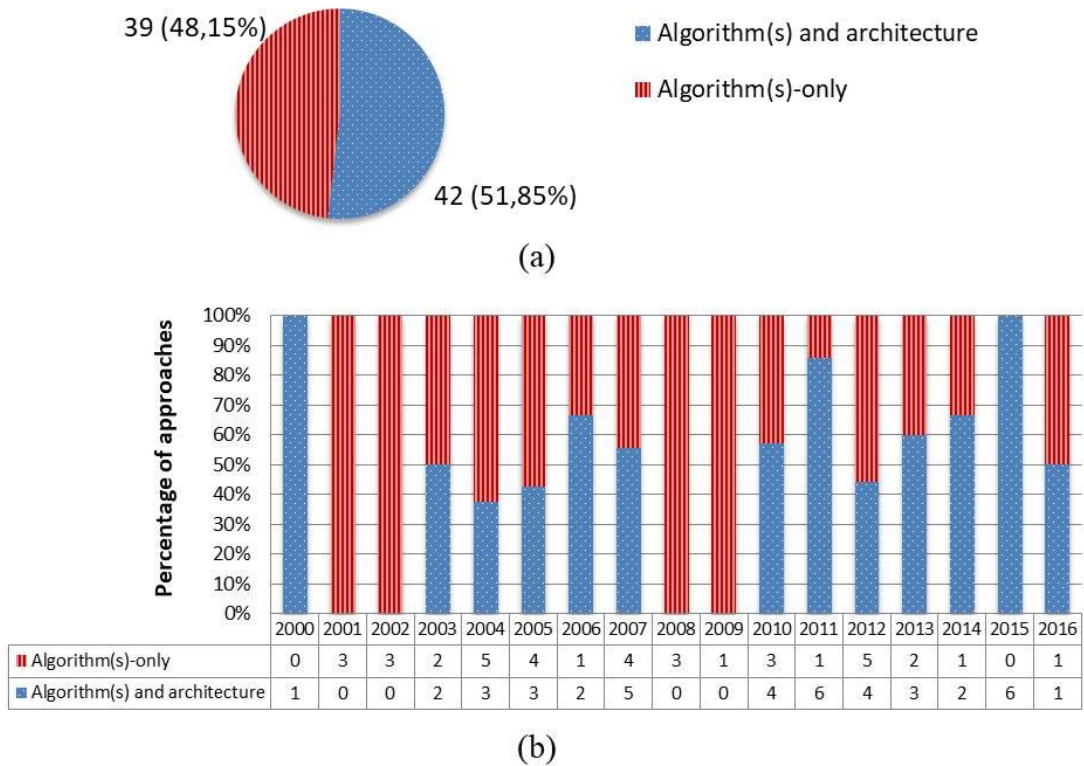

Figure 9: Number and percentage of approaches per type of contribution: (a) total, (b) over the years

*RQ3.2 - How generic are the solutions presented?*

For answering this question, we have classified approaches' solutions in three main types: *Problem-specific, Domain-specific* and *Generic*. *Problem-specific* solutions correspond to approaches that try to solve a specific problem in a specific domain, e.g., an algorithm for adapting the path of mobile sensors in order to improve monitoring precision when supervising water quality. *Domain-specific* solutions are considered for approaches supporting adaptive monitoring in a specific domain but without constraining the solution to a specific problem, e.g., an approach for supporting monitoring rules adaptation in WS-BPEL processes through dynamic weaving. Finally, the *Generic* category corresponds to solutions that can be applied in any domain, e.g., a threshold-based solution for changing monitoring systems' sampling rate.

We have conducted a first cycle coding (see Section 3.3), and as a result, we have found 64 approaches proposing *Problem-specific* solutions, 14 providing *Domain-specific* solutions and 3 presenting *Generic* ones (see Fig. 10a). Fig. 10b shows the percentage of approaches per type of solution over the years. As it can be noticed, most of the *Domain-specific* solutions belong to approaches with contributions published after 2007.

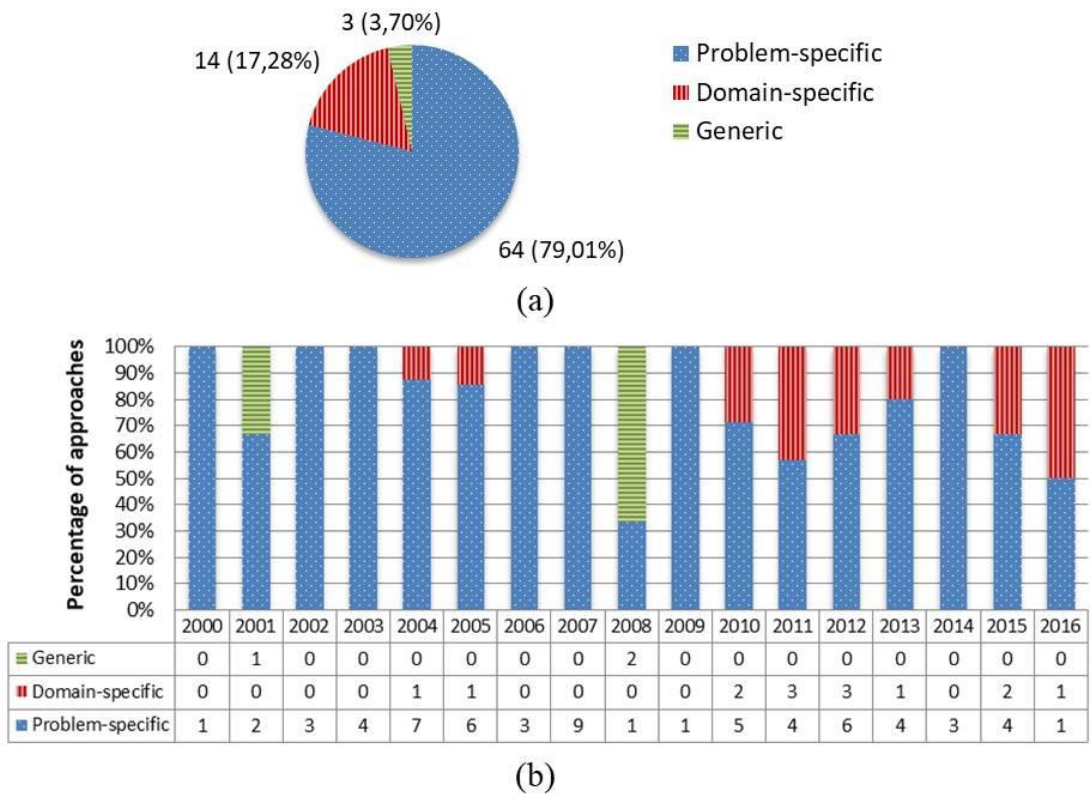

Figure 10: Number and percentage of approaches per type of solution: (a) total, (b) over the years

### 4.4. RQ4. How adaptive monitoring is conducted by the approaches?

*RQ4.1 - What is the purpose of adaptation?*

To answer this RQ, we have first derived from the approaches all the different adaptation purposes in the form of descriptive codes, i.e., inductive first cycle coding (see Section 3.3). Then, we have classified these purposes into different types. Fig. 11a shows the number of approaches motivated by the different types of purposes. The most popular type is *Solve a trade-off* (42 approaches). There are some approaches motivated by two types of purposes; however, except for one pair of purposes that was used by two approaches (*Provide adaptation capabilities* and *Respond to changes*), each combination of purposes was used just by one approach. In Fig. 11b, the percentage of approaches per type of purpose is displayed by year. This figure shows that *Solve a trade-off* has motivated approaches for a long timespan (from 2001 to 2016).

The average amount of approaches per type of purpose is 14,17. As it can be noticed, *Solve a trade-off* type of purpose is by far above this average, thus we further explore it. This type of purpose is composed of 14 different trade-offs, here we focus in the most relevant ones, i.e., trade-offs motivating more than one approach. From the most to the least popular, we find: *Improve the understanding about the SuM* while *Reducing the overhead associated with monitoring* (14 approaches), *Improve the understanding about the SuM* while *Reducing the energy consumption* (9 approaches), *Improve monitoring data accuracy* while *Reducing the overhead associated with monitoring* (6 approaches), *Improve monitoring data accuracy* while *Not exceeding available resources* (2 approaches) and *Improve monitoring coverage* while *Reducing energy consumption* (2 approaches).

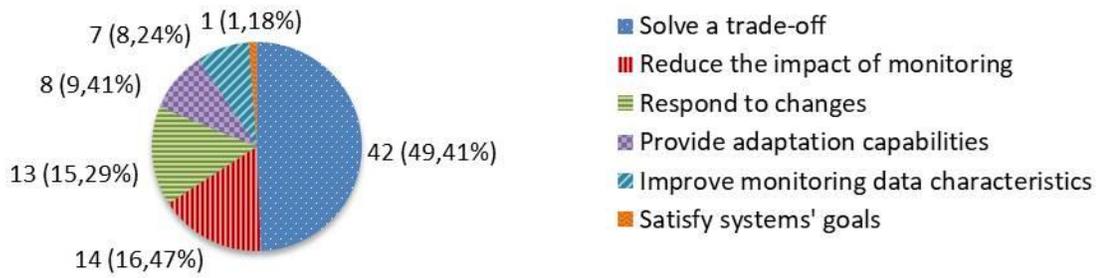
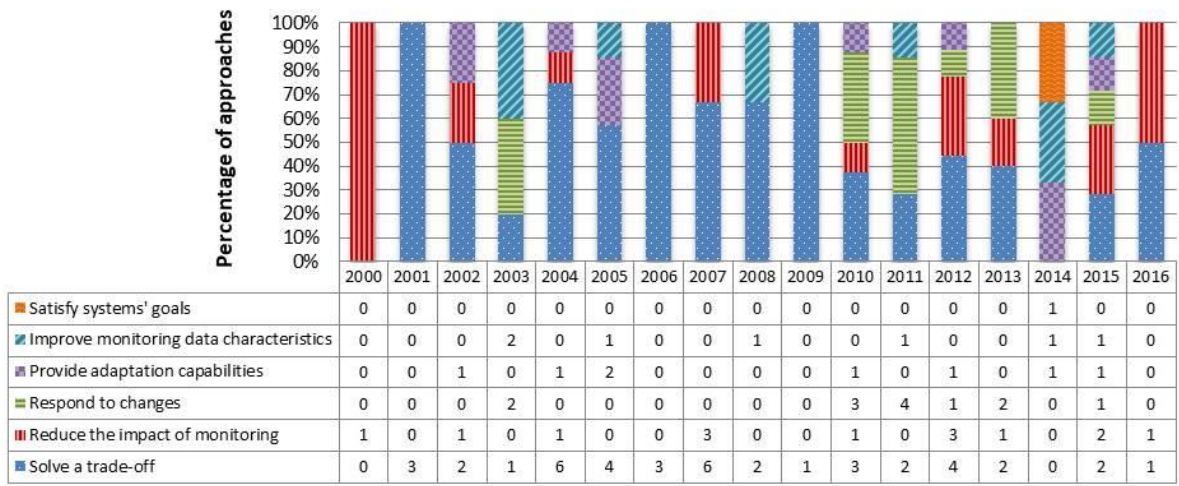

Figure 11: Number and percentage of approaches
per type of adaptation purpose: (a) total, (b) over the years

*RQ4.2 - What is adapted?*

In order to address this question, we have derived codes that describe elements adapted by existing approaches during the data extraction process, i.e., we have conducted inductive first cycle coding (see Section 3.3). Fig. 12a shows the elements that existing approaches adapt and the number of approaches that support the adaptation of each element. In Fig. 12b, we provide the percentage of approaches per year that support the adaptation of a specific element. As it can be noticed, the most adapted elements are the *Sampling points* (37 approaches) and the *Sampling rate* (25 approaches). Moreover, the relevance the adaptation of these elements over the years is evident, particularly for the *Sampling points* (present from 2000 to 2016 except for 2008).

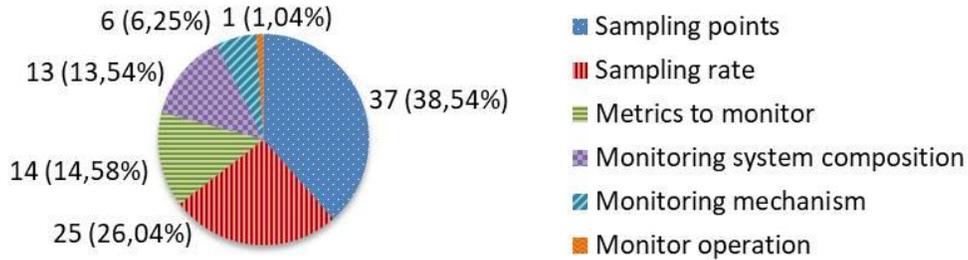
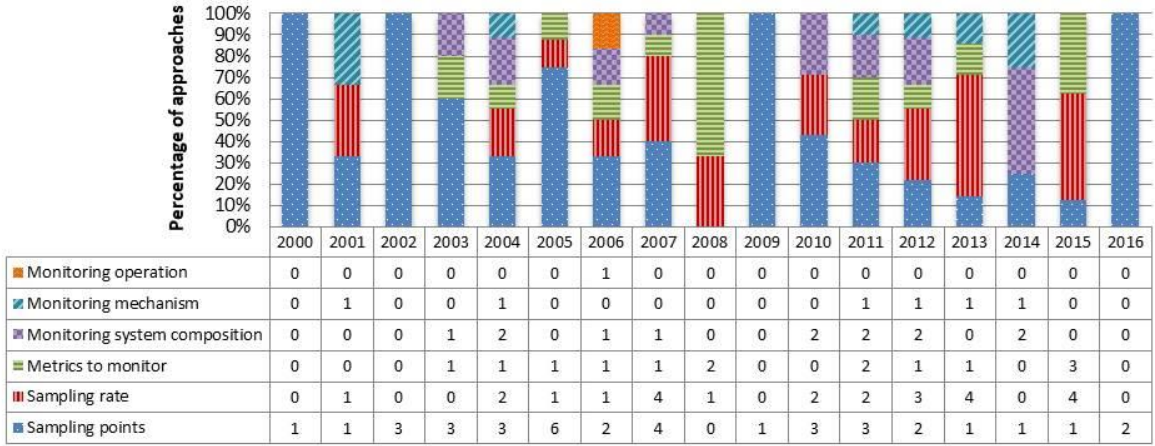

Figure 12: Number and percentage of approaches per element adapted: (a) total, (b) over the years

Some of the approaches support the adaptation of more than one element. From the most to the least popular, the most relevant combinations of elements supported by existing approaches, i.e., combinations supported by more than one approach, are: *Metrics to monitor* and *Sampling points* (4 approaches) and *Metrics to monitor* and *Sampling rate* (2 approaches).

*RQ4.3 - What triggers adaptation?*

For answering this question, we have applied both cycles of coding (Section 3.3). First, we have derived a set of codes for describing the different triggers we have found in existing approaches. Then, we have grouped them by type. In Fig. 13a, the number of approaches per type of trigger is presented while Fig. 13b shows the percentage of approaches per trigger type over the years. A *Suspected problem* is the most common factor that triggers adaptation in existing approaches (28 approaches); the relevance of this type of trigger is corroborated by its long and continuous presence in approaches over the years (from 2001 to 2016 with just two years of absence, 2002 and 2014). Some of the approaches use more than one type of factor for triggering the monitoring adaptation process. The most relevant combinations of types of triggers we have found, i.e., combinations used by more than one approach, are: *Suspected problem* and *Time* (2 approaches) and *SuM or monitoring system changes* and *Monitoring requirements changes* (2 approaches).

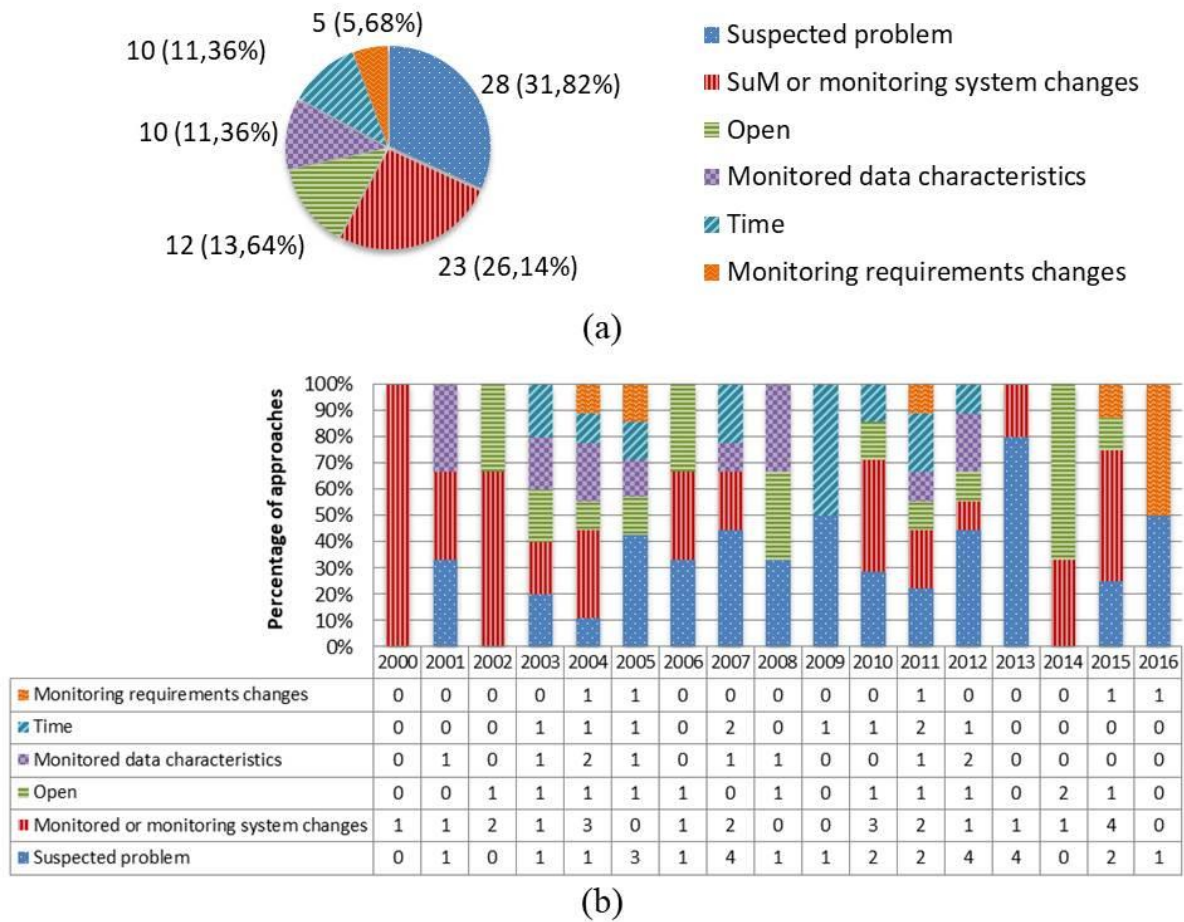

Figure 13: Number and percentage of approaches
per type of adaptation trigger: (a) total, (b) over the years

According to data shown in Fig. 13a, the average amount of approaches per type of trigger is 14,67. Thus, we further explore the *Suspected problem* and *SuM or monitoring system changes* types. *Suspect problem* type of trigger is composed of 7 triggers, the most relevant, i.e., triggers present in more than one approach, from the most to the least popular, are: *Monitoring system component anomaly* (11 approaches), *Requirement or constraint violation* (5 approaches), *Requirement or constraint likely to be violated* (4 approaches), *SuM component anomaly* (3 approaches), *SuM component likely to present an anomaly* (2 approaches) and *Likely environmental problem* (2 approaches). Anomalies in systems' components include faults. On the other hand, the *SuM or monitoring system changes* category is composed of 4 triggers, from most to less popular: *SuM state changes* (16 approaches), *SuM components de/activation* (4 approaches), *Monitoring system components addition/removal* (2 approaches) and *Execution context changes* (2 approaches). Thus, in conclusion a change in the SuM state is the most popular trigger.

*RQ4.4 - How is analysis performed?*

To answer this question, we have conducted an inductive first cycle coding for identifying analysis solutions, and then a second cycle coding for grouping them by type. Categories for grouping approaches that do not perform analysis or do not provide details about how it is performed have also been created. Fig. 14a shows the categories created as well as the number of approaches per category. As it can be noticed, most of the approaches use a specially designed *Algorithm* for conducting the analysis task (28 approaches) followed by solutions that use *Probability/Statistics* (22 approaches).

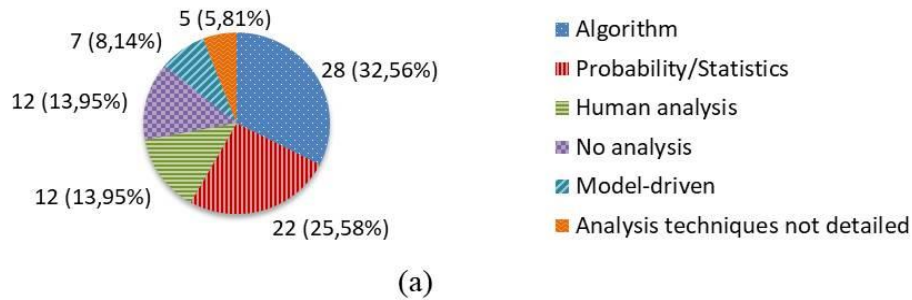

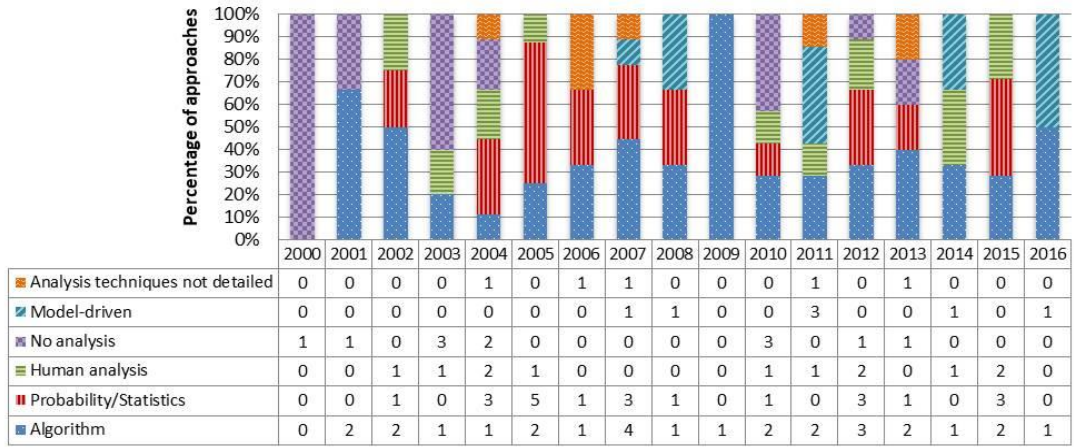

Figure 14: Number and percentage of approaches per analysis type: (a) total, (b) over the years

Fig. 14b provides the details about the percentage of approaches using specific types of analysis per year. In this figure, the relevance of the *Algorithm* category is corroborated since this type of analysis is present every year from 2001 to 2016. During the data extraction process, we have found that this type of analysis is combined with *Probability/Statistics* by two approaches. Other combinations, e.g., *Human analysis* and *Probability/Statistics*, have been also identified; however, since they were used only by one approach each, they have not been considered relevant for the purposes of this systematic mapping (i.e., finding trends). For the same reason, we have no further decomposed the most relevant categories (i.e., *Algorithm* and *Probability/Statistics*); every analysis solution in these categories is unique which do not provide information relevant for finding trends.

*R4.5 - How adaptation decisions are made?*

For addressing this question, we have derived codes based on the type of criterion used by existing approaches for making adaptation decisions, i.e., inductive first cycle coding (see Section 3.3). Resulting codes are shown in Fig. 15a. *Polices* is the most used type of criterion for conducting the decision-making process in existing approaches (49 approaches). In Fig. 15b, we provide an overview of the percentage of approaches using the different types of decision-making criteria over the years. This figure show clearly that *Policies* have played an important role in decision-making processes since, apart from being the most used type of criterion, they have been utilized by approaches since 2000 till 2016 (except for 2009). *Policies* have also been combined in existing approaches with the other types of decision-making criteria. Concretely, 4 approaches have combined them with *Human decision*, 3 with *Rules* and 2 with an *Objective function*. We have not found any combination that does not involve *Policies*.

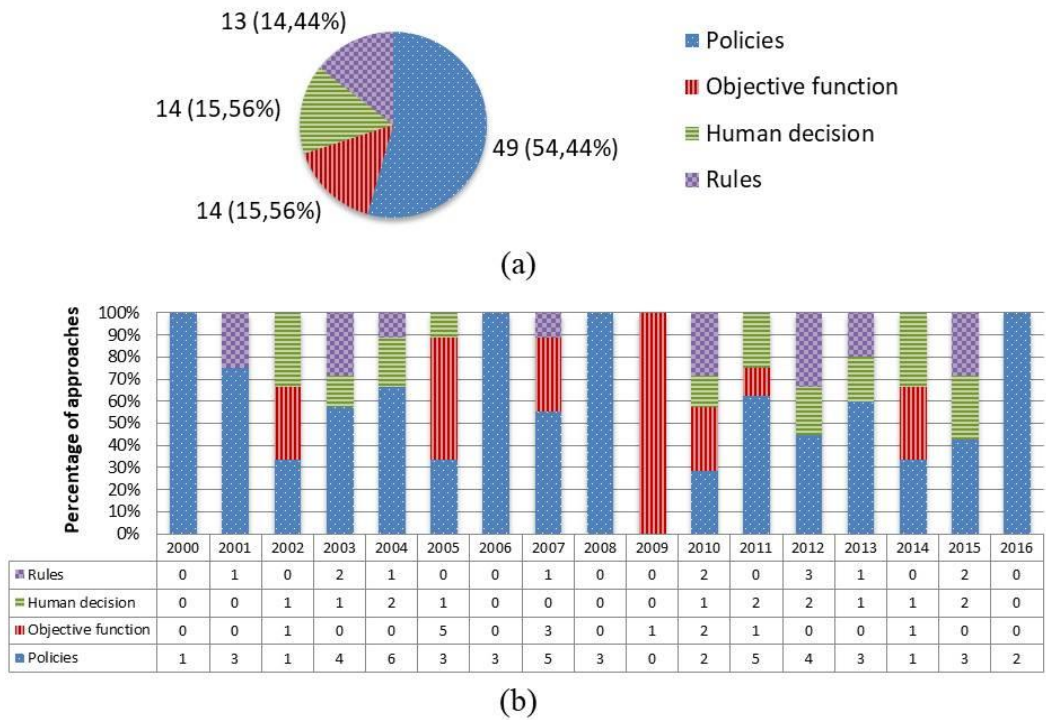

Figure 15: Number and percentage of approaches per decision-making type: (a) total, (b) over the years

*RQ4.6 - How adaptation decisions are enacted in the monitoring system?*

Three codes for describing the type of enactment process have been derived from existing approaches in order to answer this question: *Automatic*, *Semi-automatic* and *Manual*. Fig. 16a shows the distribution of approaches among the different types, resulting from a first cycle coding (see Section 3.3). The percentage of approaches using the different types of enactment per year is shown in Fig. 16b. *Automatic* is by far the type of enactment most used by existing approaches (70 approaches published between 2000 and 2016). During the data extraction, we have identified four approaches that support both *Automatic* and *Manual* enactment.

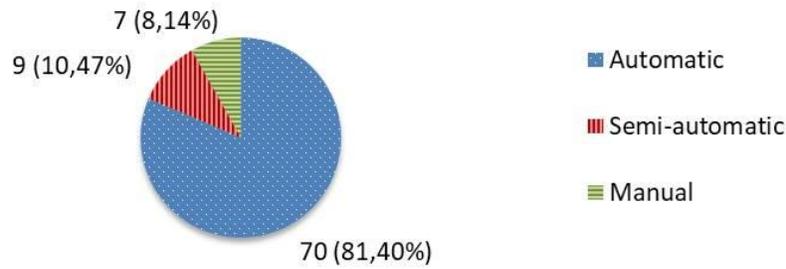

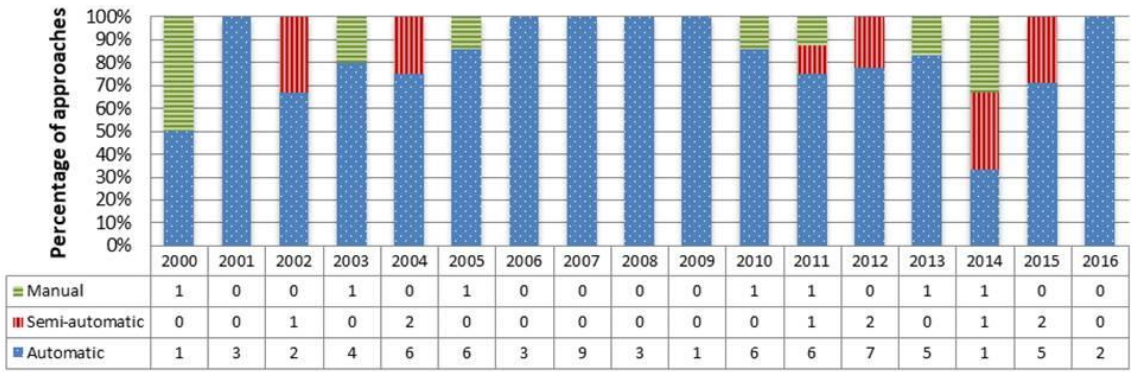

Figure 16: Number and percentage of approaches per enactment type: (a) total, (b) over the years

*RQ4.7 - What type of adaptation is executed?*

For addressing this RQ, we have considered two codes that describe two different types of adaptation: *Structural* and *Parameter*. The first one refers to changes in the structure of the monitoring system, such as the exchange of components or a new composition of components [27]. The second one refers to changes in the monitoring system's parameters, such as the change of the sampling rate or the change of the list of metrics to monitor [27]. Using these codes we have conducted a first cycle coding (see Section 3.3). Fig. 17a shows that in existing approaches most of the adaptation decisions have been translated into *Structural* monitoring systems' changes (45 approaches). From the extracted data we have identified three approaches that support both types of adaptations. In Fig. 17b, we provide an overview of the percentage of approaches per type of adaptation over the years.

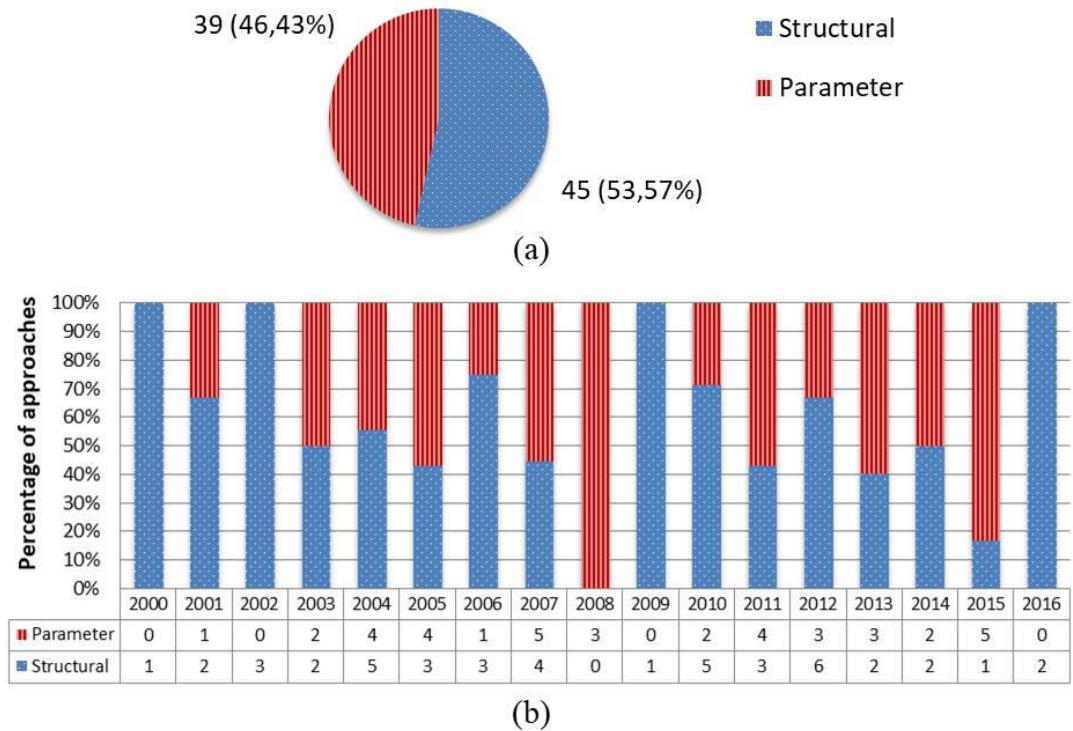

Figure 17: Number and percentage of approaches
per type of adaptation executed: (a) total, (b) over the years

### 4.5. RQ5. How adaptive monitoring approaches are evaluated?

*RQ5.1 - What type of evaluation is performed?*

To address this question, we have derived a set of codes based on the types of evaluation we have found in existing approaches, if any, and conducted a first cycle coding (see Section 3.3). The resulting codes are: *Experimentation*, *Industry use case* and *No evaluation*. In Fig. 18a, we provide the information about the number of approaches per type of evaluation (approaches presenting theoretical examples where grouped into the *No evaluation* category). *Experimentation* has been the most used method for evaluating existing approaches (59 approaches). Information about how the different types of evaluation have been used over the years by existing approaches is displayed in Fig. 18b. *Experimentation* and *Industry use case* have been present over long timespans (2000 to 2016 for *Experimentation* and 2001 to 2016 for *Industry use case*). We have not found approaches utilizing more than one type of evaluation.

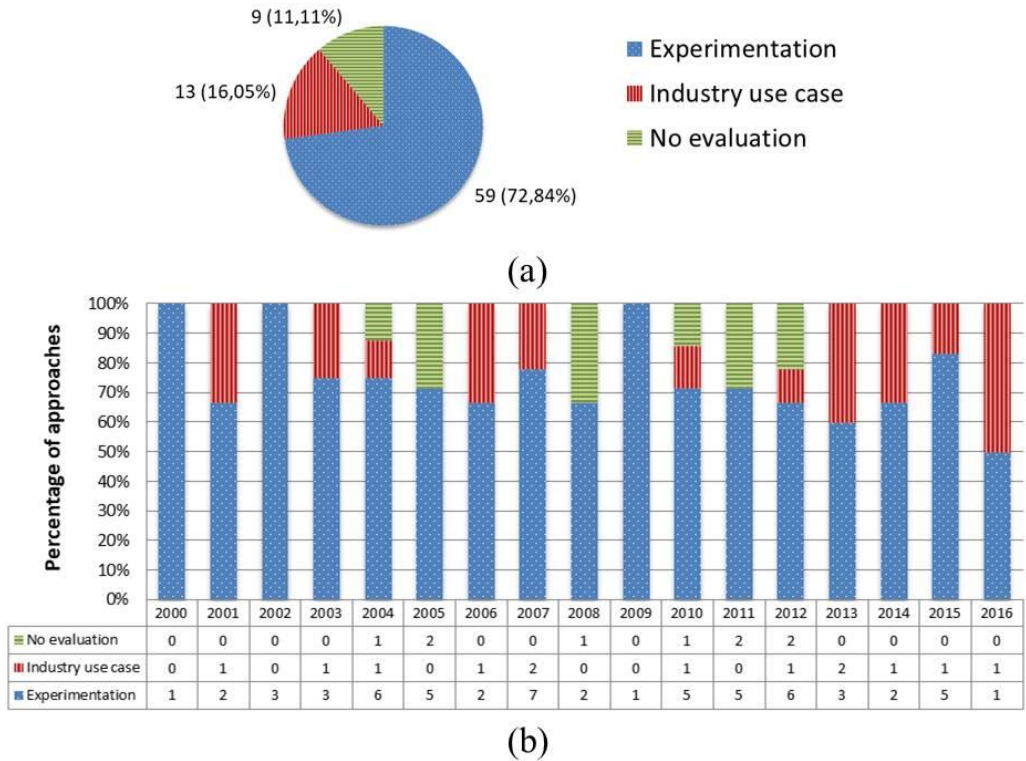

Figure 18: Number and percentage of approaches per evaluation type: (a) total, (b) over the years

*RQ5.2 - In which type of systems is the evaluation performed?*

For answering this question, codes describing the different types of systems in which approaches are evaluated have been progressively added during the data extraction process, i.e., we have applied inductive first cycle coding (see Section 3.3). Fig. 19a shows the number of approaches per system type (approaches not evaluated have been grouped into a *No evaluation* code). The most common types are: *Sensor networks* (composed of non-mobile sensors) present in 18 approaches, *Service/component-based systems* utilized by 17 approaches and *Networks* used for evaluating 14 approaches. Fig. 19b shows the percentage of approaches evaluated in a specific type of system per year. According to this figure, most of the approaches evaluated in *Sensor networks* were published before 2008, while a wave of approaches performing evaluation in *Service/component-based systems* has been experienced after 2009. Evaluation in *Networks* cannot be characterized based on this figure. We have identified only one approach that has been evaluated in more than one type of system (*Sensor networks* and *Clouds/Grids*).

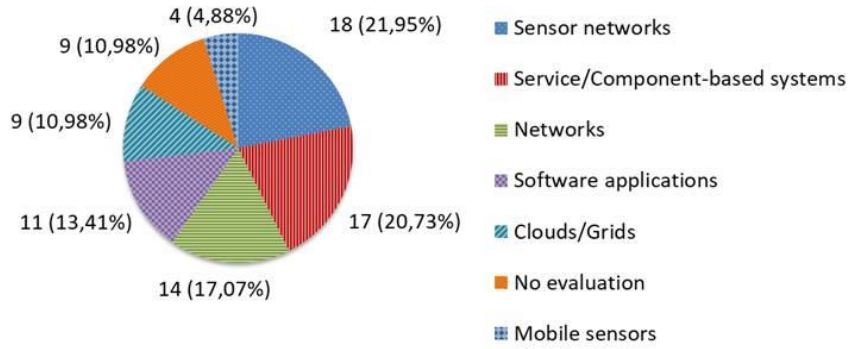

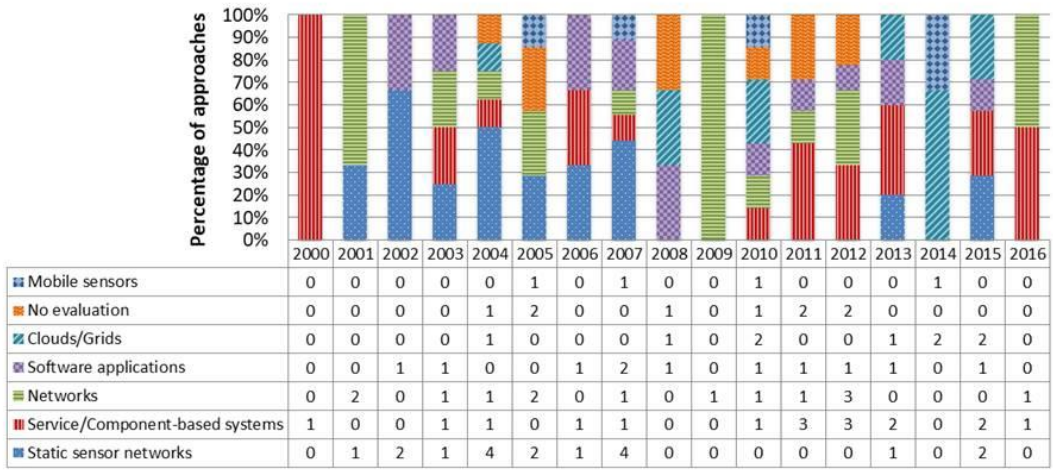

Figure 19: Number and percentage of approaches per type of system in which the evaluation is performed: (a) total, (b) over the years

## 5. Discussion

In this section, we apply data mining techniques to the resulting codes of this section in order to find further insights about the current state of the art of adaptive monitoring approaches. We are interested on identifying patterns in the approaches that cannot be easily determined by traditional analysis techniques, such the ones used in this section. Moreover, we analyze the results and discuss our findings for each research question.

### 5.1. Data mining

Data mining refers to the process of applying machine learning algorithms to data sets in order to discover patterns within the data. It is useful when human analysis is not feasible (e.g., very large amounts of data or high-dimensional data) and/or patterns are non-obvious. In literature reviews, data mining has been applied, for instance, in the form of text mining for supporting the study selection process [28–31]. In this work, we use data mining techniques for identifying patterns in the demographic characteristics of existing approaches (RQ2), the ways they present and conduct adaptive monitoring (RQ3, RQ4) as well as the evaluation processes (RQ5).

In order to perform the data mining analysis, we have defined a set of variables based on the codes extracted when addressing the RQs (see Section 4). The complete list of variables used is provided in Appendix B (Table B1). As we have mentioned in previous sections, for conducting this analysis we have used the data mining tool Weka and the rule-based algorithm called JRip. Concretely, we have run a classification for each variable, using the variable in turn as the class attribute of that run. All resulting classifiers were evaluated using stratified 10-fold cross validation. The performance metrics produced in Weka that are averaged using the cross validation for each classifier include precision, recall and f-measure [11]. From each run, we have tabulated the classifier's resulting rules and the values of the performance metrics mentioned before.

In Appendix B (see Fig. B1) we provide an overview of the performance metrics' values obtained for the classifier of each variable. The closer the values are to 1.0, the better the performance of the classifier. The criteria for deciding whether a classifier is good enough depend on the specific use case. In our case, we have not precedents for establishing criteria since we have not found any other review applying data mining to RQs answers for finding patterns. Thus, we have decided to consider classifiers with precision, recall and f-measure greater than or equal to 0.9. As a result, classifiers for 17 out of the 47 analyzed variables were considered relevant. In Appendix B (Table B2), the list of rules that compose these 17 relevant classifiers is provided. In the rest of this section, we discuss the results we have obtained in Section 4 and complement the analysis with the relevant patterns we have found in this sub-section.

### 5.2. Analysis of results

*RQ1. What is adaptive monitoring?*

The diversity of research fields from which studies of our systematic mapping have emerged, has certainly contributed to the diversification of the vocabulary used for referring to adaptive monitoring. This phenomenon can be clearly seen in Fig. 2, in which we have presented the different terms categories utilized by 81,81% of the studies as alternatives to the term "adaptive monitoring". Most of these terms are domain-specific and in consequence cannot be reutilized in all the research fields.

One of the objectives of this study is to find a generic definition for the term "adaptive monitoring". In Section 4.1, we have looked for definitions in the studies of our systematic mapping. As a result, we have found only one definition. Moreover, this definition is not complete and generic enough for applying to the different realizations of adaptive monitoring we have found in this review, e.g., monitoring systems composition adaptation based on SuM state changes could be not covered with this definition.

In order to achieve our objective, we have adjusted the definition proposed by Moui et al. [25,26] and created a generic definition for the term "adaptive monitoring":

> "Adaptive monitoring is the ability a monitoring system has to modify its structure and/or behavior in order to respond to internal and external stimuli such changes in their execution context, functional and non-functional requirements, systems under monitoring or the monitoring system itself"

In this definition, all monitoring systems are treated equally (sensor networks, component-based software monitoring systems, instrumentation systems, etc.) which is beneficial for later standardizing other concepts applicable to all high-level monitoring systems as well, e.g., monitoring frequency adaptation. Moreover, unlike the definition proposed by Moui et al. [25,26] which only consider the adjustment of behavioral aspects, in this definition adaptation is understood as changes in the monitoring system behavior as well as in its structure. Finally, the possible triggers of the adaptation process are not constrained in our definition as they are in the Moui et al.'s.

*RQ2. What are the demographic characteristics of the studies about adaptive monitoring?*

The adaptation of monitoring systems is a lively research area with studies published every year from 2000 to 2016 (see Fig. 3). However, it is remarkable that, if we take 3-year windows, the last period (2014-2016) is the one with fewer contributions (excluding the first period 2000-2002, when the topic was formulated). Interpretation of these trends needs always to be careful. On the one hand, the third period with fewer contributions was 2008-2010 but only a 2-year shift (2010-2012) yields to the most populated window. On the other hand, the advent of domains like IoT, smart vehicles, etc., where self-adaptation and in relation to it, adaptive monitoring, is crucial, it may be expected a growth of contributions.

In terms of venue, most of the published papers in this topic have appeared in conference proceedings (see Fig. 4); the percentage is very close to the average of 25.9% reported by Ameller et al. [32] from a sample of 14 systematic mappings in software engineering. On the other hand, we have not found any dominant venue in any of the categories. For instance, the conference with more publications is the International Conference on Network and Service Management (CNSM) with 4 out of the 68 conference papers. In our opinion, this situation is due to the diversity of research fields in which adaptive monitoring is present. That is, papers are mainly published in venues specialized in the research field they belong to. This fact contributes to the isolation of solutions per research field and in some cases even per research communities. The need of venues in which the adaptive monitoring topic is central per se and research from different fields could be found and compared would help to promote this area.

From the type of publication perspective (see Fig. 5), it is not surprising that the majority of the papers are from *Academy* (academics usually are more motivated to submit papers to conferences and journals [22]); however, the number of papers from *Industry* indicates that adaptive monitoring is also a topic of interest of practitioners as well. Moreover, this interest has been present almost every year considered in this review (see Fig. 5b).

In this RQ, we have also explored how studies are distributed geographically. With not surprise, we have found that authors of North American and European organizations are the most active researchers in the area (see Fig. 6). However, from 2010 publications from North American organization have dramatically decreased, 82,46% of their studies have been published before 2010 (see Fig. 6b). The opposite effect has happened to European organizations' amount of publications. From this observation, we expect more European contributions in the next years in this topic than North American. Even though, *USA* which is the main contributor in *North America* is by far the country with most published papers (see Fig. 7). Although this is the usual situation in Computing Science as reported by Ruiz [33], the difference is much greater (26.4% of the publications are from USA in this report). We do not expect others countries to reach the same amount of publications in the short-term.

Finally, in this RQ we have analyzed how studies are organized in approaches. As a result, we have identified many different approaches (see Fig. 8). The most prominent ones (approaches with more contributions) are from the *Networks monitoring* and the *Monitoring systems* (in general) research fields. While studies composing the approach for *Networks monitoring* do not have interaction with studies of other fields (derived from citation data), the studies of the approach for supporting adaptation of *Monitoring systems* (in general) interact with studies of *Networks monitoring* as well as *Service-based systems monitoring* fields. In general, we have found that studies do not interact with studies of others research fields. As mentioned before, the interaction among research fields' proposals is beneficial for many reasons, such finding synergies and opportunities, normalizing challenges, etc. Regarding the application of data mining in the codes of this question, no relevant classifiers have been found.

*RQ3. What is proposed by the approaches?*

The distribution of approaches among the two types of contributions identified is quite even (see Fig. 9). However, it can be noticed in Fig. 9b that approaches that present algorithms supported by architectural proposals have been more and more proposed in the last years. The opposite happens to approaches contributing with only algorithms. Regarding the generalizability level of the solutions (see Fig. 10), the majority is *Problem-specific* and cannot be reutilized or extended for dealing with other issues or supporting other adaptation functionalities. However, this type of solutions in some cases could be aggregated for instance, in order to solve a group of problems in a specific domain.

A concrete example of aggregation could be an approach that combines the context-aware e-health monitoring proposal of Mshali et al. [3] with a re-configurable service-based monitoring system infrastructure (e.g., Villegas et al. [34]) for supporting an energy-efficient monitoring system that can incorporate new sensors at runtime. Regarding the few *Generic* solutions we have found, unfortunately, they are all algorithmic solutions, not complete enough for providing a unified software engineering solution to current adaptive monitoring systems of the different research fields. Finally, when applying data mining to the codes derived in this RQ no relevant classifiers have been found.

*RQ4. How adaptive monitoring is conducted by the approaches?*

Unlike in previous RQs, relevant classifiers have been found for at least one code of each research sub-question of this RQ (see Table B2 in Appendix B). For instance, for adaptation purposes, we have found that *Satisfying system's goals* is a purpose very unlikely to find in existing approaches (pattern that can also be visualized in Fig. 11), and that *Solve a trade-off* purpose do not usually motivate approaches in conjunction with other purposes (pattern expectable since in case of appearing in the same approach, purposes would form part of the trade-off). Apart from the patterns, in Fig. 11b, we have noticed that after 2009 the variety of purposes considered by approaches has increased. In the next years, we expect a behavior similar to this since we consider that given the diversity of software systems, users' needs, and execution contexts, more and more varied factors motivating adaptive monitoring will emerge.

Regarding what is adapted, a classifier confirming what is shown in Fig. 12, regarding the unlikeliness of finding an approach adapting the monitor operation, has resulted. We have also found a classifier that positively relates *Monitoring system composition* adaptation to *Structural* changes and negatively the last one with *Sampling rate* adaptation. This makes sense since one can expect *Structural* changes when re-composition is required and a *Parameter* changes when the adaptation is of a variable such the *Sampling rate*. For the adaptation triggers, we have found a classifier that corroborates that *SuM or monitoring system changes* type of trigger is not combined with other types apart from *Monitoring requirements changes*. Another classifier relates positively *Open* triggers with *Human analysis* which is reasonable considering that in most of the approaches *Open* triggers represent humans making the decision, by any reason, of triggering the adaptation process. From the chronological data shown in Fig. 12b and Fig. 13b, we have not further found relevant information.

A strong positive relation has been found between *Human analysis* and *Human decision* codes (two classifiers relate them; see Table B2 in Appendix B); particularly, in cases when adaptations are executed manually. Regarding the decision-making criteria, a second classifier for *Policies* has resulted, the pattern indicates that in general approaches do not combine *Policies* with *Objective functions* or *Rules* for making decision and that in most of the cases decisions made using *Policies* are execute automatically. Specifically, the relation of *Policies* with *Automatic* enactment makes sense since systems' owner usually design policies for being evaluated and executed automatically and in this way reduce the need of human intervention. Finally, we would like to remark that in general, analysis solutions are developed in an ad-hoc manner while decision-making methods are reutilized by different approaches.

For the types of enactment, we have found a classifier that positively relates *Manual* enactment with *Human decision* criteria, a second one that negatively relates *Human analysis* with *Automatic* enactment and a third one that indicates that *Semi-automatic* enactment is not usually supported together with the other two types. Given the relations we have found, we can say that approaches in which the adaptation process is started by humans tend to position the whole process in a human-driven manner (i.e., analysis, decision-making and enactment are not automatized). Regarding the chronological information (see Fig. 16b), it can be noticed that the involvement of humans in the adaptation process has been retaken after 2009. This is aligned with the need identified by Cheng et al. [35] and ratified by Krupitzer et al. [27] of considering the users in the adaptation process to ensure their trustiness. However, this participation should be kept as less intrusive as possible so that the automatic adaptation process performance is not affected. Finally, for *Structural* and *Parameter* adaptation we have found that in general they are not both supported by a single approach, i.e., there is a negative relation between them.

*RQ5. How adaptive monitoring approaches are evaluated?*

The distribution of the usage of the different evaluation types over the years is almost the same for all the types. Regarding the types of systems in which these evaluations take place, we have found a classifier that indicates that approaches evaluated in *Mobile sensors* are usually found in academic papers and use objective functions as decision-making criterion. Furthermore, those approaches do not tend to trigger adaptations periodically but instead use specific triggers. Particularly, the relation between the type of system and the decision-making criterion makes sense since approaches for *Mobile sensors* in many cases try to solve a trade-off (e.g., distance traveled vs phenomenon understanding) which is usually translated into an objective function. From the chronological perspective, in Fig. 19b, it can be noticed that *Sensor networks* popularity has dramatically decreased after 2007 (83,33% of approaches using *Sensor networks* for evaluating their solutions have been published from 2000 to 2007).

The last classifier that has resulted from the data mining analysis (see Table B2 in Appendix B) does not represent any new insight regarding our understanding about how approaches are evaluated. Instead, it confirms that our coding mechanism is correct, i.e., we expected to find a *No evaluation* code for the type of system in all the approaches coded with *No evaluation* code as type of evaluation. This kind of rules could be beneficial during the data extraction process for checking the correctness of codes assignation and reducing the probability of misunderstandings. Thus, the benefits of applying data mining to qualitative analysis codes are twofold, finding hidden relations and checking the correctness of the method's implementation.

## 6. Conclusions

This work has presented a systematic mapping study on adaptive monitoring focused on the adaptation of the elements directly related to the data gathering activity. The study aims at giving a comprehensive overview of the current state of the art of the adaptive monitoring topic and improving the understanding about how approaches from different research fields (tend to) conduct the adaptation process. For this purpose, we have followed a systematic review protocol that has allowed us to identify 110 studies organized in 81 proposals for supporting adaptive monitoring in a variety of research fields. The studies have been used for addressing a series of research questions we have defined as part of the review process. The analysis has been thorough, relying on coding and data mining for a deep understanding of the answers to the research questions. We consider that the results we have obtained in this review can be useful in the standardization of adaptive monitoring concepts (e.g., utilizing the codes we have developed for describing the different elements), as well as in the development of more complete, flexible, reusable and generic software engineering solutions for supporting adaptive monitoring in a variety of systems. For instance, proposing solutions that ensure the support of all types of adaptations or all types of elements' adaptation.

From the results obtained, we have also realized that future approaches may consider the abstraction of problem-specific techniques such analysis techniques which in most of the cases are ad-hoc algorithms (same applies for ad-hoc algorithms proposed for actually solving objective-functions in the decision-making phase). Solutions that allow systems' owners to experiment with different techniques are highly desirable. Moreover, the use of standardized reference models such the MAPE-K loop [36,37], widely used in the domain of self-adaptive systems, for placing the different activities carried out during the adaptation process, will ease the comparison and reusability between proposals. As future work, we would like to encourage researchers to analyze in more detail the analysis and decision-making techniques of the approaches we have presented in this study. For instance, systematic literature reviews can be conducted for evaluating different aspects of existing techniques (such performance, correctness, applicability, etc.).

From our side, we plan to propose a software engineering solution that satisfies the requirements listed previously in this section. Our objective is to start utilizing terms for the adaptation process applicable to any type of adaptive monitoring system, as well as to define a reusable architecture for supporting coordinately normal monitors' operation and their adaptation process. The idea is to develop a solution applicable to new a legacy monitoring systems through the separation of generic and system-specific functionalities. We also plan to develop a framework implementing the ideas of our adaptive monitoring proposal in order to facilitate the systematic development of adaptive monitors and evaluate it in different use cases involving different types of monitoring systems.

**Acknowledgements**

Thanks to CONACYT and I2T2, for the PhD scholarship granted to Edith Zavala. This work is partially supported by the Spanish project GENESIS (TIN2016-79269-R).

**Appendix A. Systematic Mapping References**

**Table A1**

## Appendix B. Data Mining Variables and Results

**Table B1**
Variables used in data mining

| Research sub-question | Id | Variable | Values |
|---|---|---|---|
| RQ2.1 | $v_1$ | Year *(of last approach contribution)* | 2000-2016 |
| RQ2.2 | $v_2$ | Type of publication *(of last approach contribution)* | Conference, Journal, Workshop |
| RQ2.3 | $v_3$ | Type of paper *(of last approach* | Industry, Academy |
| RQ2.4 | $v_4$ | Continent *(of last approach contribution)* | North America, South America, Europe, Asia, Oceania |
| RQ3.1 | $v_5$ | Type of contribution | Algorithm(s) and architecture, Algorithm(s)-only |
| RQ3.2 | $v_6$ | Solution generalizability | Problem-specific, Domain-specific, Generic |
| RQ4.1 | $v_7$ | Improve monitoring data characteristics | True, False |
| RQ4.1 | $v_8$ | Provide adaptation capabilities | True, False |
| RQ4.1 | $v_9$ | Reduce the impact of monitoring | True, False |
| RQ4.1 | $v_{10}$ | Respond to changes | True, False |
| RQ4.1 | $v_{11}$ | Satisfy systems' goals | True, False |
| RQ4.1 | $v_{12}$ | Solve a trade-off | True, False |
| RQ4.2 | $v_{13}$ | Metrics to monitor | True, False |
| RQ4.2 | $v_{14}$ | Monitoring operation | True, False |
| RQ4.2 | $v_{15}$ | Monitoring mechanism | True, False |
| RQ4.2 | $v_{16}$ | Monitoring system composition | True, False |
| RQ4.2 | $v_{17}$ | Sampling points | True, False |
| RQ4.2 | $v_{18}$ | Sampling rate | True, False |
| RQ4.3 | $v_{19}$ | Suspected problem | True, False |
| RQ4.3 | $v_{20}$ | SuM or monitoring system changes | True, False |
| RQ4.3 | $v_{21}$ | Monitored data characteristics | True, False |
| RQ4.3 | $v_{22}$ | Monitoring requirements changes | True, False |
| RQ4.3 | $v_{23}$ | Time | True, False |
| RQ4.3 | $v_{24}$ | Trigger open | True, False |
| RQ4.4 | $v_{25}$ | Algorithm | True, False |
| RQ4.4 | $v_{26}$ | Model-driven | True, False |
| RQ4.4 | $v_{27}$ | Analysis techniques not detailed | True, False |
| RQ4.4 | $v_{28}$ | Human analysis | True, False |
| RQ4.4 | $v_{29}$ | No analysis | True, False |
| RQ4.4 | $v_{30}$ | Probability/Statistics | True, False |
| RQ4.5 | $v_{31}$ | Human decision | True, False |
| RQ4.5 | $v_{32}$ | Objective function | True, False |
| RQ4.5 | $v_{33}$ | Policies | True, False |
| RQ4.5 | $v_{34}$ | Rules | True, False |
| RQ4.6 | $v_{35}$ | Manual | True, False |
| RQ4.6 | $v_{36}$ | Automatic | True, False |
| RQ4.6 | $v_{37}$ | Semi-automatic | True, False |
| RQ4.7 | $v_{38}$ | Parameter | True, False |
| RQ4.7 | $v_{39}$ | Structural | True, False |
| RQ5.1 | $v_{40}$ | Type of evaluation | Experiment, Industry use case, No evaluation |
| RQ5.2 | $v_{41}$ | Software applications | True, False |
| RQ5.2 | $v_{42}$ | Clouds/Grids | True, False |
| RQ5.2 | $v_{43}$ | Mobile sensors | True, False |
| RQ5.2 | $v_{44}$ | Network | True, False |
| RQ5.2 | $v_{45}$ | No evaluation | True, False |
| RQ5.2 | $v_{46}$ | Sensor networks | True, False |
| RQ5.2 | $v_{47}$ | Service/Component-based systems | True, False |

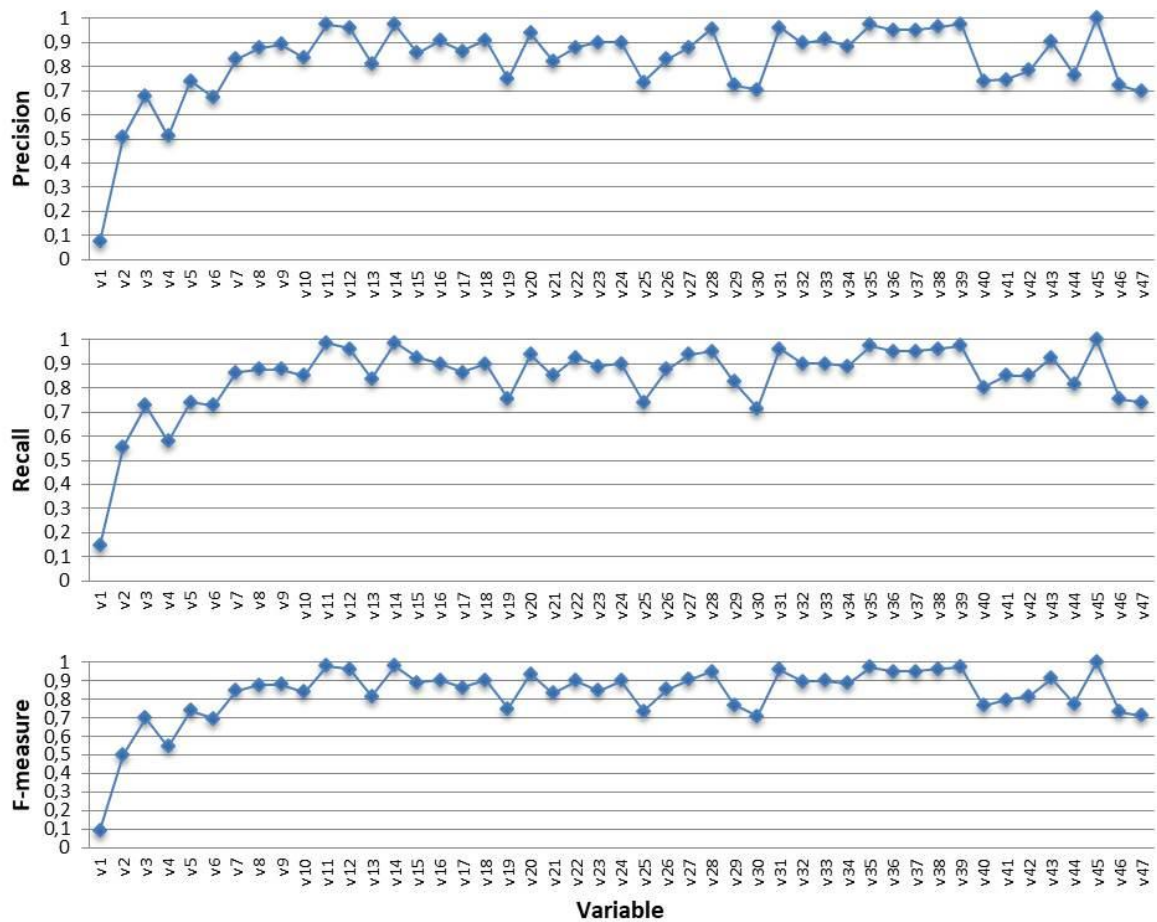

Figure B1: Resulting *precision*, *recall* and *f-measure* per variable classifier

**Table B2**
Resulting relevant data mining classifiers

| Variable | Rules | Interpretation |
|---|---|---|
| **Satisfy** | *(for all)* Satisfy system's goals = False | In general, approaches are not motivated by the |

| systems' goals | | purpose of satisfying system's goals. |
|---|---|---|
| Solve a trade-off | *If* ((Reduce the impact of monitoring = True) or (Respond to changes = True) or (Improve monitoring data characteristics = True) or (Provide adaptation capabilities = True)) <br>   *then* (Solve a trade-off = False) *(True otherwise)* | Some adaptation purposes (reduce the impact of monitoring, respond to changes, improve monitoring data characteristics or provide adaptation capabilities) do not usually motivate an approach in conjunction with solving a trade-off purpose. |
| Monitoring operation | *(for all)* Monitoring operation = False | In general, approaches do not aim at adapting the monitoring operation. |
| Monitoring system composition | *If* (Structural = True and Sampling points = False and Suspected problem = False) <br>   *then* (Monitoring system composition = True) *(False otherwise)* | Structural changes executed on monitoring systems by approaches are usually done for enacting monitoring system composition adaptation decisions, as long as they do not correspond to sampling points' adaptations and the adaptation trigger is not a suspected problem. |
| Sampling rate | *If* (Structural = False and Sampling points = False) <br>   *then* (Sampling rate = True) *(False otherwise)* | Parameter changes are usually executed by approaches for adapting the sampling rate, except in the cases of non-structural sampling points' adaptation. |
| SuM or monitoring system changes | *If* (Suspected problem = False and Trigger open = False and Monitored data characteristics = False and Time = False) <br>   *then* SuM or monitoring system change = True *(False otherwise)* | Approaches triggering adaptation by SuM or monitoring system changes do not tend to consider some kinds of triggers (suspected problem, open trigger, monitored data characteristics and time). |
| Trigger open | *If* (Human analysis = True and SuM or monitoring system change = False) <br>   *then* (Trigger open = True) *(False otherwise)* | Approaches considering human analysis that do not trigger adaptations by SuM or monitoring systems changes, tend to leave the adaptation trigger open. |
| Human analysis | *If* (Human decision = True) <br>   *then* (Human analysis = True) *(False otherwise)* | Approaches considering human-based decision-making usually also consider human-based analysis. |
| Human decision | *If* (Human analysis = True or Manual = True) <br>   *then* (Human decision = True) *(False otherwise)* | Approaches considering human analysis or manual enactment of the adaptation decisions tend to conduct decision-making supported by humans. |
| Policies | *If* (Objective function = True or Automatic = False or Rules = True) <br>   *then* (Policies = False) *(True otherwise)* | Policies are mainly used by existing approaches for making adaptation decisions, except for approaches that do not support automatic enactment or use objective functions or rules as decision-making criteria. |
| Manual | *If* (Human decision = True and Semi-automatic = False) <br>   *then* (Manual = True) *(False otherwise)* | Approaches considering human-driven decision-making process tend to enact adaptations semi-automatically or manually. |
| Automatic | *If* (Human analysis = True) <br>   *then* (Automatic = False) *(True otherwise)* | Most of the approaches considering human analysis do not consider automatic enactment. |
| Semi- | *If* (Automatic = False and Manual = False) | Approaches supporting semi-automatic |
| **automatic** | ***then*** (Semi-automatic = True) *(False otherwise)* | enactment do not support other kinds of enactment. |
| **Parameter** | ***If*** (Structural = False) <br> ***then*** (Parameter = True) *(False otherwise)* | In general, approaches do not support the execution of both types of adaptation in a single solution. |
| **Structural** | ***If*** (Parameter = True and Monitoring system composition = False) <br> ***then*** (Structural = False) *(True otherwise)* | |
| **Mobile sensors** | ***If*** (Objective function = True and Type of paper = Academy and Time = False) <br> ***then*** (Mobile sensors = True) *(False otherwise)* | Approaches evaluated in mobile sensors systems do not trigger adaptations periodically and use objective functions for conducting their decision-making process. Moreover, most of them have been published by academics. |
| **No evaluation** | ***If*** (Type of evaluation= No evaluation) <br> ***then*** (No evaluation= True) *(False otherwise)* | Approaches we have grouped in the *No evaluation* category in RQ5.1 were also correctly classified in RQ5.2 as not evaluated. |